\documentclass[twocolumn,trackchanges]{aastex631}
\newcommand{\prot}{$P_{\rm rot}$}
\newcommand{\tauc}{$\tau_{\rm cz}$}
\newcommand{\ro}{Ro}
\newcommand{\vsini}{$v\sin i$}
\newcommand{\bprp}{$G_{\rm BP}-G_{\rm RP}$}
\newcommand{\rvar}{$R_{\rm var}$}

\begin{document}

\title{Slowly Rotating Fully Convective Stars at $\lesssim$ 1 Gyr: Evidence from the Hyades and Praesepe}

\newcommand{\mdot}{$M_{\odot}$}
\newcommand{\osu}{Department of Astronomy, The Ohio State University, Columbus, 140 W 18th Ave, OH 43210, USA}
\newcommand{\ccapp}{Center for Cosmology and Astroparticle Physics (CCAPP), The Ohio State University, 191 W. Woodruff Ave., Columbus, OH 43210, USA}

\author[0000-0003-4769-3273]{Yuxi(Lucy) Lu}
\affiliation{\osu}
\affiliation{\ccapp}

\author[0000-0002-7549-7766]{Marc H. Pinsonneault}
\affiliation{\osu}
\affiliation{\ccapp}

\author{Jos\'e G. Fern\'andez-Trincado}
\affiliation{Centro de investigaci\'on en Astronom\'ia, Facultad de Ingenier\'ia, Ciencia y Tecnolog\'ia, Universidad Bernardo O'Higgins, jose.fernandez@ubo.cl, Av. Viel 1497, Santiago, 8370993, Chile}

\author{Richard R. Lane}
\affiliation{Centro de investigaci\'on en Astronom\'ia, Facultad de Ingenier\'ia, Ciencia y Tecnolog\'ia, Universidad Bernardo O'Higgins, jose.fernandez@ubo.cl, Av. Viel 1497, Santiago, 8370993, Chile}

\author[0000-0002-1379-4204]{Alexandre Roman-Lopes}
\affiliation{Departamento de Astronomía, Facultad de Ciencias, Universidad de La Serena,
Av. Raúl Bitrán 1305, La Serena, Chile}

\author[0000-0002-3481-9052]{Keivan G.\ Stassun}
\affiliation{Department of Physics and Astronomy, Vanderbilt University, Nashville, TN 37235, USA}

\author[0000-0002-1533-9029]{Emily K.\ Pass}
\affiliation{Kavli Institute for Astrophysics and Space Research, Massachusetts Institute of Technology, Cambridge, MA 02139, USA}

\author[0000-0003-1479-3059]{Guy S.\ Stringfellow}
\affiliation{University of Colorado Boulder, Boulder, CO 80309 USA}



\begin{abstract}
We report the discovery of 10 stars in the Praesepe and Hyades clusters with unusually slow rotation that are likely not field contamination.
Even though our search was general, all slow rotators are in the fully convective domain.
We find that the fraction of slow rotators in the clusters is more than $5\sigma$ above the distribution expected from field contamination.
They provide a potential link between the period distribution of the young open clusters and the field. 
These stars could be a distinct birth population, or they could be an indication that the onset of strong magnetic breaking is earlier than previously realized in this domain. 
Photometric variability and X-ray luminosity suggest the later but more data is needed to confirm.
Searches for similar populations in younger clusters will help determine when slow rotators first emerge and constrain the timescale and physical mechanisms governing angular-momentum loss in fully convective stars. 
The discovery of these stars also highlights the importance of long-baseline photometric observations for detecting the long rotation periods of low-mass stars.
\end{abstract}



\section{Introduction} \label{sec:intro}
The use of stellar rotation as an age indicator depends on whether stars eventually lose their ``memory'' of their initial rotation rates; that is, whether their subsequent rotational evolution erases most of the dependence on their initial conditions.
Stars are born with a broad distribution of rotation periods, shaped by their initial angular momentum and early star--disk interactions \citep{Terebey1984, Koenigl1991, Irwin2009, Rebull2018}.
Magnetic braking subsequently removes angular momentum and causes stars to spin down.
Since the torques for rapid rotators are larger than those for slow rotators, convergence in rotation rates at late times naturally occurs. 
Once this convergence occurs, rotation becomes primarily a function of stellar mass and age, providing the physical basis for gyrochronology \citep[e.g.,][]{Skumanich1972, Barnes2003, Epstein2014}.
If a subset of stars are born with extremely slow rotation, their weaker spin-down torques will cause them to converge more slowly toward a common rotational sequence, as initially rapid rotators require time to spin down and catch up. 
These stars can therefore retain substantial memory of their initial conditions, allowing stars of the same mass and age to have very different rotation periods and preventing rotation from providing a unique age.

The predicted convergence of rotation rates is clearly seen in GK stars \citep[see for example the data collated in ][]{GodoyRivera2021}. 
However, the pattern is less clear for the lowest mass stars, where data is sparse in older clusters.
Open-cluster studies have revealed a characteristic ``waterfall'' in the rotation-period distribution near the fully convective boundary: rotation periods increase toward the boundary and then drop abruptly immediately beyond it, accompanied by saturated magnetic activity \citep[e.g.,][]{Irwin2007, Nunez2015, Douglas2016, Douglas2017, Rebull2018, Douglas2019, Ilin2019, Kiman2021}. 
Empirically, the angular-momentum distribution of young stars in open clusters appears to evolve more smoothly and monotonically with stellar mass and age than does the corresponding rotation-period distribution \citep{Somers2017, Kounkel2022,Kounkel2023}. 
In particular, \citet{Kounkel2022} showed that expressing rotation in terms of angular momentum produces a comparatively continuous empirical sequence from FGK through M stars over ages of $\sim$10--1000 Myr.
This distinction is important because the observed rotation period reflects both the angular momentum of the star and its mass- and age-dependent moment of
inertia.
Thus, complex structure in the period--mass distribution does not necessarily imply similarly complex underlying angular-momentum evolution.

In contrast to the rapidly rotating M dwarfs observed in open clusters, field-star surveys reveal slow rotators across a broad range of masses \citep[e.g.,][]{Newton2016, Lu2022}, including fully convective stars whose inferred ages suggest that slow rotation may be established relatively early in their evolution \citep[e.g.,][]{Pass2022, Lu2024, Chiti2024}.
However, because ages are difficult to infer for the field stars, the existence of slowly rotators in the field alone does not reveal how rapidly stellar rotation converges toward a common age--rotation relation or how much memory of the initial conditions is retained.
Identifying slowly rotating stars in coeval young clusters is therefore crucial for distinguishing between these possibilities and determining when the initial rotation distribution is sufficiently erased for gyrochronology to become a reliable clock.

One significant concern is that young slow rotators may have been missed by previous surveys because of selection effects.
FC stars are intrinsically faint, and their long rotation periods (many $>35$ days) require long temporal baselines to measure reliably.
Most rotation periods for open clusters have been measured using K2 \citep{K2}, whose $\sim$70-day observing campaigns and broad optical bandpasses limit sensitivity to slowly rotating cool M dwarfs, or TESS \citep{tess}, whose scattered-light systematics and relatively short observing baselines make rotation periods longer than $\sim15$ days increasingly difficult to recover \citep[e.g.,][]{Holcomb2022, Colman2024, Hattori2025, Boyle2026}.
Another reason slow rotators are often discounted in cluster studies is the potential for field-star contamination. 
Even a small fraction of contaminants can disproportionately populate the slow-rotating tail of a cluster’s observed rotation distribution. 
Distinguishing genuine cluster slow rotators from field contaminants and observational incompleteness is therefore essential for constraining the true rotational evolution of young stars.

The Zwicky Transient Facility \citep[ZTF;][]{ztfdata, ztftime} has been observing the northern sky for more than eight years, and its relatively red optical bandpasses make it well suited for measuring the long rotation periods of cool K and M dwarfs. 
The long temporal baseline is particularly important because the uncertainty of individual photometric measurements can be comparable to the variability amplitude. 
Repeated sampling of the same rotation cycle can substantially increase the effective signal-to-noise ratio; for example, a 60-day period is sampled over more than 40 cycles across an eight-year baseline. 
This repeated coverage therefore improves our ability to detect rotation signals in faint, slowly rotating stars.
In this paper, we address this missing part of the cluster rotation distribution by searching for long-period rotators in the Hyades and Praesepe using ZTF. 
We report the discovery of slowly rotating (\prot$> 35$ days) FC stars in both clusters and demonstrate that at least a small population of these stars are consistent with cluster membership rather than being field contaminants.
These objects provide cluster counterparts to the slowly rotating fully convective population previously identified in the field, substantially reducing the apparent discrepancy between field and cluster rotation distributions. 
If confirmed as cluster members, their existence demonstrates that some FC stars can reach rotation periods longer than 35 days by ages $\lesssim1$ Gyr, either through substantial early angular momentum loss or because some stars are born with relatively slow rotation.
In contrast, G- and K-dwarfs are expected to reach rotation periods of $\sim35$ days only at ages $\gtrsim8$ Gyr \citep{Lu2025}.

In \autoref{sec:data_method}, we describe the data and rotation-period measurements. 
In \autoref{sec:result}, we present the newly identified population of slow rotators. 
We first investigate whether these stars could be field-star contaminants in \autoref{subsec:FieldCont}. 
We then examine their rotation-period distributions and magnetic activity properties in \autoref{subsec:periodDist} and \autoref{subsec:activityRotation}, respectively. 
Finally, we discuss the implications of this population for stellar dynamo theory and gyrochronology of fully convective stars in \autoref{sec:dynamo}.

\section{Data \& Method} \label{sec:data_method}
\subsection{Open cluster membership \& existing rotation period catalogs}\label{subsec:opencluster}
We chose the Hyades and Praesepe not only because their proximity enables more reliable M-dwarf membership determination and facilitates checks for field contamination, but also because their relatively young ages of $\sim700$ Myr make them prime targets for searching for young, slowly rotating M dwarfs. 
In addition, both clusters were observed by K2, providing space-based light curves that can be compared directly with the ground-based ZTF observations.
The cluster membership list is taken from \cite{Hunt2023}, in which they applied Hierarchical Density-Based Spatial Clustering of Applications with Noise (HDBSCAN) algorithm to recover cluster members using 5D kinematic information (proper motions, parallax, and Galactic longitudinal and latitude) from Gaia DR3 \citep{Gaiadr3}.
A detailed analysis of cluster membership and field contamination is discussed in \autoref{subsec:FieldCont}.

The literature rotation catalogs for the Hyades and Praesepe are taken from \cite{vanlane2025}, which combines rotation catalogs from multiple literature sources based on K2 \citep{K2} and Transiting Exoplanet Survey Satellite (TESS; \citealt{tess}) observations \citep{Radick1987, Radick1995, Hartman2010, Delorme2011, Douglas2016, Douglas2019, Rebull2017, Rampalli2021, GodoyRivera2021, Long2023}.

In addition, we compare the cluster rotation periods with rotation period measurements for field stars.
The field-star sample is compiled from several literature sources \citep{McQuillan2014, Santos2019, Lu2022, Lu2024_gyro, Boyle2026}, combining measurements from the main Kepler mission \citep{kepler}, TESS, and the ZTF \citep{ztfdata, ztftime}.

\subsection{Light curves \& other rotation period catalogs}
We perform our analysis using the $r$-band light curves, which provide the best temporal sampling among the ZTF filters while maintaining greater spot contrast.

We download light curves from Zubercal \citep[ZTF Ubercal;][]{drake2025zubercal}\footnote{Available at \url{http://atua.caltech.edu/ZTF/Zubercal.html}.}.
Zubercal is a newly recalibrated photometric data set that improves upon the standard ZTF PSF photometry by simultaneously calibrating all PSF measurements obtained within a given detector quadrant (1/64 of the full field of view) and filter during a single observing night. 
Compared to the standard IPAC-calibrated ZTF photometry distributed through IRSA, which contains complex systematic photometric residuals of up to $\sim5$\%, Zubercal removes these systematics and delivers substantially improved photometric precision.

To identify potential failure modes of the Lomb--Scargle (LS) periodogram arising from irregular temporal sampling \citep[see also \autoref{subsec:protmeasure};][]{VanderPlas2018, Dungee2022}, we additionally analyzed all available K2 self-flat fielding (K2SFF) light curves downloaded using \texttt{lightkurve} \citep{lightkurve} and generated TESS light curves from Full Frame Images (FFIs) using \texttt{unpopular} \citep{unpopular}. 
The K2SFF light curves \citep{Vanderburg2014} correct for the pointing-dependent systematics introduced by spacecraft motion while preserving long-timescale astrophysical variability, while \texttt{unpopular} employs a causal pixel model to remove instrumental systematics while preserving long-period signals.
For each of the Zubercal light curves, we bin the observations using a 0.1-day window and remove outliers that deviate by more than 3$\sigma$ from the main trend. 
For the K2SFF and TESS light curves, we remove linear trends from each observing season, bin the light curves using a 3-hour window, and remove outliers that deviate by more than 3$\sigma$ from the main trend.
We search for rotation periods with LS periodogram in the range of 0.2-40 days for K2 light curves and 0.2-20 days for TESS light curves, matching close to their observing season length.

In addition, we also compared our measured rotation periods with those reported in large Kepler and TESS rotation-period catalogs \citep{McQuillan2014, Santos2019, Boyle2026}.

\subsection{Measuring known rotation periods with ZTF}\label{subsec:protmeasure}
\begin{figure*}[ht!]
\centering
\includegraphics[width=\textwidth]{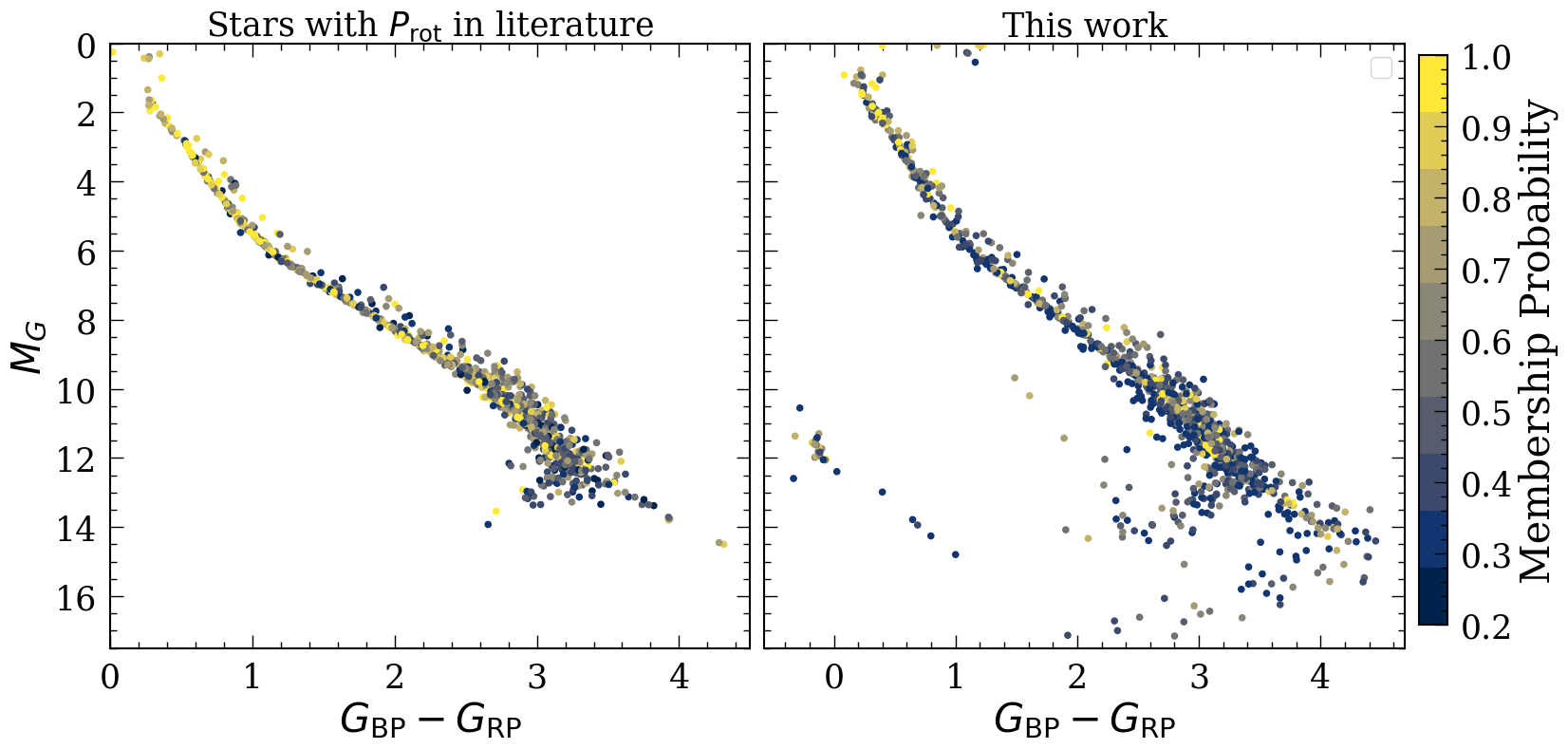}
\caption{The color–magnitude diagrams (CMDs) of Praesepe and the Hyades, colored by the membership probabilities reported by \cite{Hunt2023}. 
The left panel shows stars with publicly available period measurements \citep[886 for Praesepe and 179 for the Hyades;][]{Radick1987, Radick1995, Hartman2010, Delorme2011, Douglas2016, Douglas2019, Rebull2017, Rampalli2021, GodoyRivera2021, Long2023, vanlane2025}, while the colored points in the right panel represent cluster members for which we searched for rotation periods using Zubercal $r$-band light curves (428 for Praesepe and 761 for the Hyades).
The membership probability decreases toward lower masses, as the photometric and proper-motion measurements become increasingly uncertain due to the faintness of these stars.
This does not necessarily indicate that the lower-mass stars are non-members; rather, their fainter magnitudes make it more difficult to distinguish cluster members from field stars based on the available measurements.
\label{fig:1}}
\end{figure*}

We measured rotation periods in the Zubercal $r$-band for all stars in the cluster rotation catalogs, regardless of their membership probabilities, to enable a direct comparison with previous studies. 
Lower-mass stars typically have lower membership probabilities due to larger uncertainties in their Gaia DR3 astrometric measurements, resulting from them being fainter compared to hotter dwarfs.
We further discuss and assess the potential impact of field contamination in \autoref{subsec:FieldCont}.
This resulted in 886 and 166 stars with measured periods, and 428 and 761 additional stars for which we searched for periods from Zubercal light curves in Praesepe and Hyades, respectively.
The color magnitude diagram (CMD) combining the two clusters is shown in \autoref{fig:1}.
The points are colored by their membership probability from \cite{Hunt2023}.
The left panel shows stars with period measurements from various sources (see \autoref{subsec:opencluster}), while the right panel shows stars without publicly available period measurements.

For each star, we downloaded the Zubercal light curve using a 1.2 arcsec search radius and excluded data points with photometric flags or magnitude uncertainties $>0.1$ mag. 
Most stars excluded due to photometric flags are affected by saturation, for which Zubercal performs poorly at magnitudes brighter than $14$ mag.
We only analyzed stars with more than 300 epochs of observations to ensure reliable period measurements.
We search for periods between 0.5 and 100 days using the \texttt{astropy.LombScargle} \citep{astropy:2013, astropy:2018, astropy2022} function with Modified Julian Date (MJD), magnitude, and magnitude error as input. 

For the initial vetting of rotation periods, we retained only period measurements with a maximum LS power $>0.1$, a ratio of maximum to median LS power $>10$ (analogous to a signal-to-noise ratio), and a light-curve baseline at least 2.5 times longer than the detected period.

The failure modes of the LS periodogram are primarily caused by uneven temporal sampling, which can produce alias periods, spectral leakage, spurious peaks, and incorrect identification of the true rotation period \citep{VanderPlas2018}.
The failure-mode periods ($P_{\rm obs}$) produced by uneven temporal sampling can be expressed as a function of the true period ($P_{\rm true}$):
\begin{equation}\label{eq:1}
P_{\rm obs} = \left| \frac{m}{P_{\rm true}} + \frac{n}{\delta P} \right|^{-1},
\end{equation}
where $m$ and $n$ are integers, and $\delta P$ represents the characteristic sampling period or the period of a systematic signal.
To determine the characteristic sampling periods, we compute the LS periodogram of the spectral window function over periods between 0.2 and 1000 days. 
The spectral window is constructed by assigning unit flux to each observation. 
We identify peaks that are 3$\sigma$ above the median LS power, yielding characteristic sampling periods of 0.33, 0.50, 1.00, 363.67, and 444.48 days.
We additionally include known periodic systematics, such as the $\sim$29.5 day lunar cycle, based on previous studies and empirical inspection of the ensemble period distribution.
For vetting, we also excluded any periods that lay within 10\% of these known systematics.

\begin{figure}[ht!]
\centering
\includegraphics[width=\columnwidth]{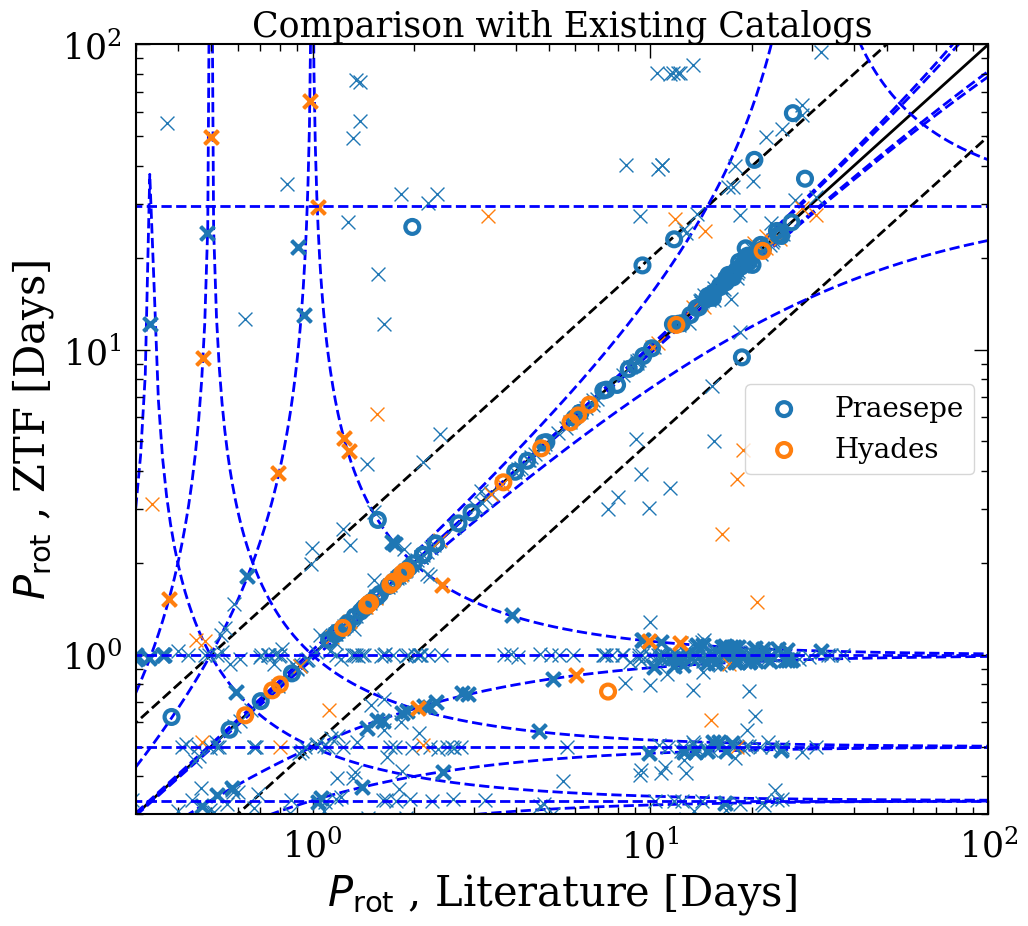}
\caption{Comparison between the literature rotation periods (see \autoref{subsec:opencluster}) and those recovered from ZTF for Praesepe (blue) and Hyades (orange). 
Circle points and crosses indicate accepted and rejected ZTF periods, respectively. 
Rejected periods correspond to stars that fail our period-quality criteria, including LS maximum power $<$0.1, LS maximum power-to-median power ratio $<$10, or periods consistent with known ground-based systematic signals (within 0.05 days of 0.5 days, within 0.1 days of 1 day, and within 0.5 days of 29.5 days).
Bold crosses highlight measurements identified as period aliases of the systematics determined from the spectral window function (0.33, 0.50, 1.00, 29.5, 363.67, and 444.48 days) following the relation described in \autoref{eq:1}. 
Aliases are identified by requiring the observed ZTF frequency, (1/$P_{\rm obs}$), to match the predicted alias frequency from the literature space-based period, (1/$P_{\rm true}$), within 10\%. 
The corresponding true periods are then recovered by applying the inverse alias relation. 
After alias correction, the ZTF periods show agreement with the literature catalogs for 93.2\% and 95.0\% of Praesepe and Hyades stars, respectively.
\label{fig:2}}
\end{figure}

\autoref{fig:2} shows a comparison between the literature periods and the periods recovered from ZTF. 
The blue points and crosses represent the accepted and rejected periods for Praesepe, respectively, while the orange symbols show the corresponding measurements for the Hyades.
The bold crosses indicate ZTF periods that lie on the expected period-alias relations described in \autoref{eq:1}.
We identify these aliases by requiring the observed ZTF frequency, $1/P_{\rm obs}$, to agree within 10\% with an alias frequency predicted from the literature period, $1/P_{\rm true}$. 
This approach is possible because the literature periods are independently measured and are not subject to the same uneven-sampling window function as the ZTF observations.
The underlying true periods can then be recovered by applying the inverse alias relation. 
After correcting for these aliases, the final agreement between the literature period catalogs and our ZTF measurements is 93.2\% and 95.0\% for Praesepe and Hyades, respectively.

\begin{figure*}[ht!]
\centering
\includegraphics[width=\textwidth]{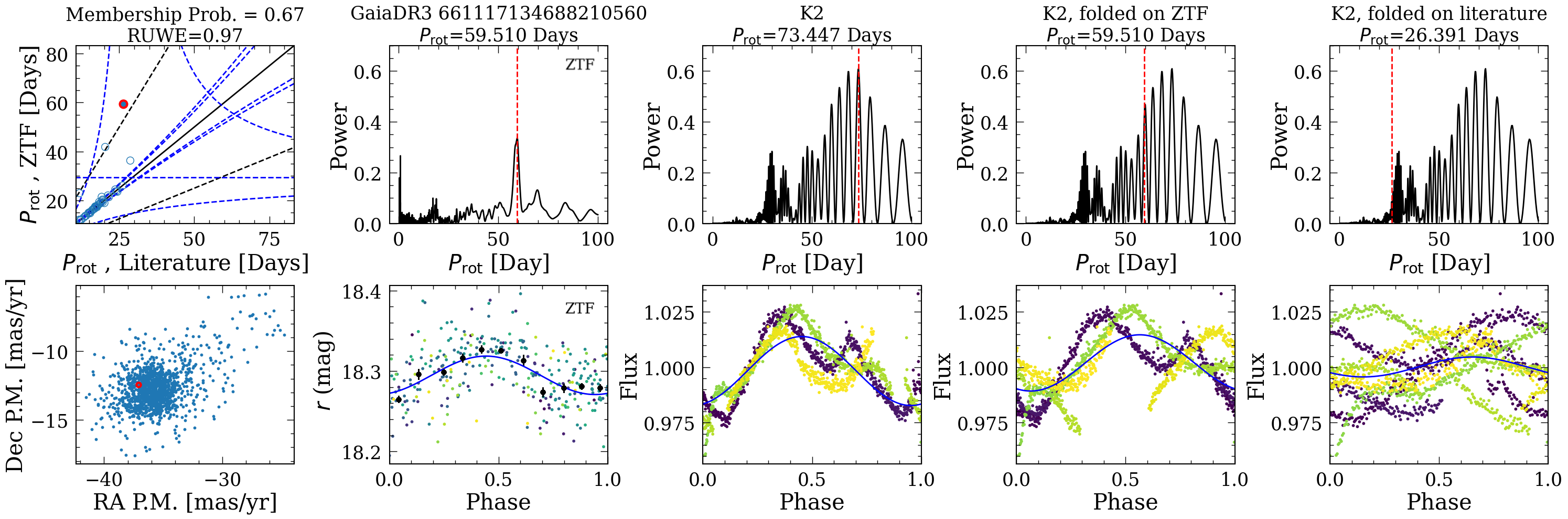}
\caption{One example from Praesepe is shown where the ZTF period is approximately twice the reported literature period.
This star is also one of the FC stars in Praesepe with a detected rotation period exceeding 35 days.
The top-left panel shows the same visualization as \autoref{fig:2} in linear space, with the targeted star highlighted by a red outline. 
The bottom-left panel shows the Gaia DR3 proper motions of the full cluster population in blue, with the targeted star again highlighted by a red outline.
The second column shows the LS power spectrum of the Zubercal light curve, highlighting the maximum power peak at the detected period, along with the corresponding folded light curve. 
Each point in the folded light curve is colored by its MJD to illustrate the temporal distribution of the observations. 
The best-fit model is shown in blue, while the median-binned folded light curve is shown as black points. 
The error bars represent the median absolute deviation of the fluxes within each bin.
The final three columns show the LS power spectrum of the K2SFF light curve and the corresponding folded light curves using the detected K2 period, the ZTF period, and the literature period reported in \cite{vanlane2025}, respectively.
The folded light curves are also colored by MJD.
While the period is less clearly identified in the K2 data due to its proximity to the observing window, the 59 day period exhibits a clear signal in the ZTF light curve. 
This demonstrates the importance of ZTF's long ($\sim$2000)-day baseline for detecting long-period rotation signals of $>$ 35 days.
\label{fig:3}}
\end{figure*}

For stars with ZTF period measurements deviating more than 10\%, which is the typical level of period variation expected from differential rotation \citep{Epstein2014}, from the reported period from the literature, we vetted the K2 and Zubercal light curves and find that most are either missing systematics (see \autoref{fig:2}, missing aliases that lay close to the alias blue dashed line), or half-period harmonics.
For example, \autoref{fig:3} shows a case where the K2 and ZTF rotation periods do not agree. 
This star is particularly interesting because the ZTF light curve exhibits a clear and well-defined peak in the LS power spectrum at a long period of $\sim$60 days, while the K2 power spectrum shows a more ambiguous signal. 
The K2 light curve contains multiple peaks around ($\sim$73) days, close to its observing window, making it difficult to identify the true rotation period. 
Although some power is present near the literature period, the short seasonal baseline of the K2 observations limits our ability to confidently determine the correct period. 
In contrast, the long temporal baseline of ZTF enables the detection of longer-period signals. 
For a 59-day rotation period, the $\sim$2000-day ZTF baseline for this star provides approximately 34 rotational cycles in the folded light curve, allowing for a more robust period measurement.

A similar case is true for Gaia DR3 661281206736808064 in Hyades, where the ZTF detected period is 41.8 days, while the reported period is 20.3 days.

\subsection{Measuring additional rotation periods with ZTF}\label{subsec:protmeasure_new}
After testing our period recovery method, we applied the same method to recover periods for stars without reported values in the literature, which include 428 stars in Praesepe and 761 stars in Hyades, of which 219 and 227 stars have more than 300 epochs in Zubercal after restricting to measurements without photometric flags and with magnitude uncertainty $<$ 0.1 mag. 
Out of these stars, 44 stars in Praesepe and 122 stars in Hyades have maximum LS power $>$ 0.1 and the ratio between the maximum and the median LS power $>$ 10.
For these stars, we downloaded and measured periods from the K2SFF and TESS light curves generated from FFIs with \texttt{unpopular}.
We also cross-matched our stars with periods from the all sky TESS catalog \citep{Boyle2026}.
For stars with measured periods from TESS, we only keep measurements with maximum LS power $>$ 0.1.

\begin{figure}[ht!]
\centering
\includegraphics[width=\columnwidth]{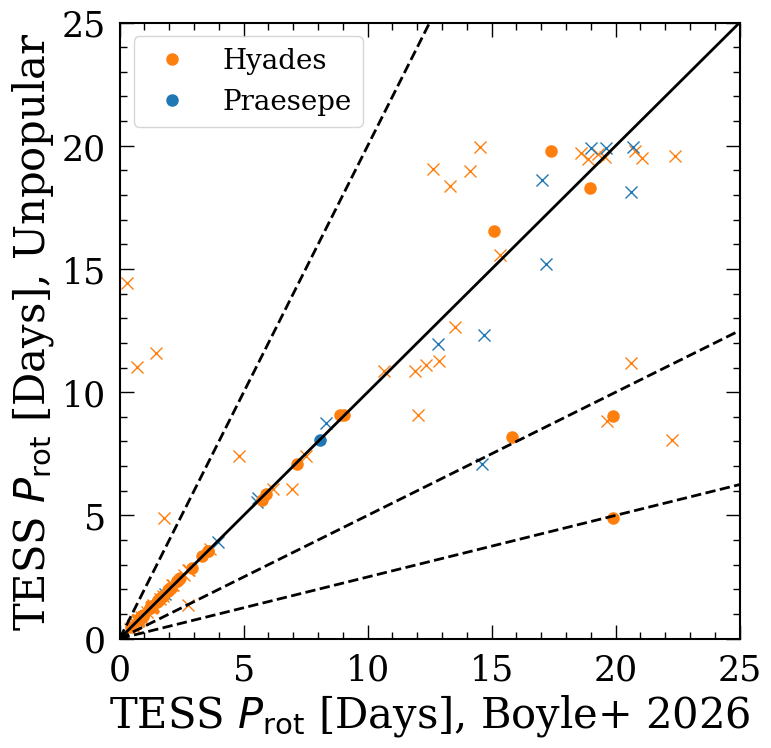}
\caption{Period measurements from TESS compared between this work and those from \cite{Boyle2026}. 
The two colors represent stars in the Hyades and Praesepe clusters as shown in the captions. 
The solid and dashed lines show the 1:1 period relation and the half- and quarter-period harmonics, respectively.
Crosses indicate stars with period measurements from our analysis but with maximum LS power $<0.1$, while circles indicate stars for which we accept the detected periods (maximum LS power $>0.1$). 
For the stars with accepted periods, our measurements agree with those from \cite{Boyle2026} within 10\% of the half-period, quarter-period, and 1:1 period relations.
10\% is chosen as it is the typical level of period variation expected from differential rotation \citep{Epstein2014}.
All periods lie along the 1:1 relation below 10 days but deviate at longer periods, highlighting the difficulty of measuring long rotation periods with TESS.
This comparison only serves as a guideline. 
All stars with accepted period measurements from ZTF are subsequently vetted by eye using all available light curves from TESS and K2.
\label{fig:4}}
\end{figure}

\autoref{fig:4} shows the comparison between the periods reported by \cite{Boyle2026} and our period measurements.
For stars with a maximum LS power $>0.1$, our measured periods agree with those from \cite{Boyle2026} to within 10\% for the half-period, quarter-period, and 1:1 period relations. 
The agreement becomes worse at periods $\gtrsim10$ days, as expected, since these periods approach the observational baseline of a single TESS observing season.
The comparison with previous period measurements serves only as a guideline for assessing the reliability of our detections. 
We later visually inspect all accepted ZTF period measurements using all available TESS and K2 light curves to ensure the robustness of the identified rotation signals.

\begin{figure*}[ht!]
\centering
\includegraphics[width=\textwidth]{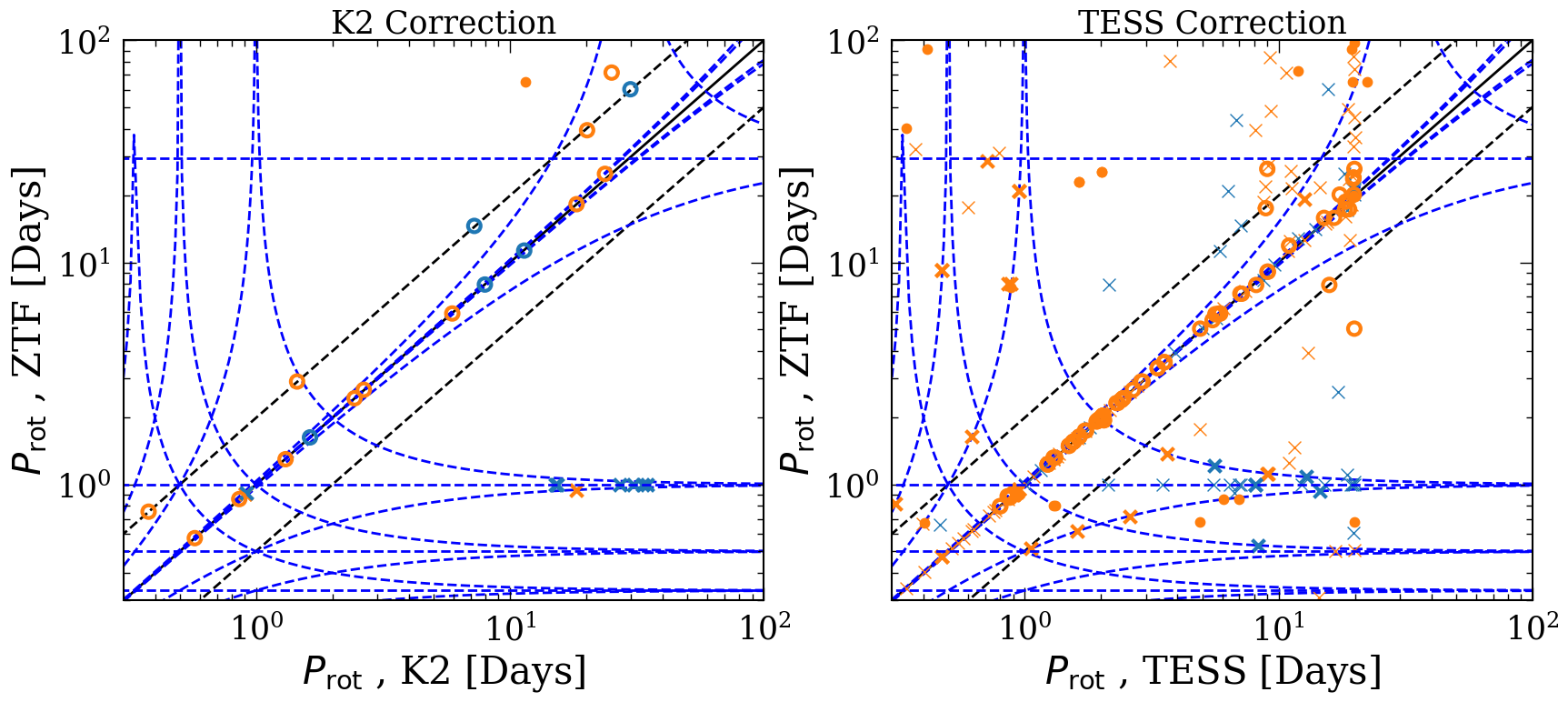}
\caption{Similar to \autoref{fig:2}, but for stars without previously reported rotation periods. 
Only stars with a ZTF maximum LS power $>0.1$ and a ratio of maximum to median LS power $>10$ are shown.
Thin crosses indicate stars whose TESS period measurements were rejected (see \autoref{fig:4}).
After vetting for systematic periods (horizontal blue dashed lines) and correcting for the failure modes (blue dashed curves), we visually inspected each star's light curve by eye.
We adopted 9 periods from TESS/K2 and excluded 6 ambiguous periods (shown as solid points).
This left us with 26 stars in Praesepe and 100 stars in the Hyades.
The empty circles indicate the subset of stars for which a K2 period is available or the maximum LS power from TESS exceeds 0.1.
\label{fig:5}}
\end{figure*}

\autoref{fig:5} shows the comparison between K2 (left) and TESS (right) with our measured ZTF period. 
The systematics identified in the spectra window (see \autoref{subsec:protmeasure} and \autoref{eq:1}) are plotted in blue dashed lines.
Of the 44 stars in Praesepe, 16 lie on the systematic lines (blue horizontal dashed lines), while 18 of the 122 stars in the Hyades do so. 
In addition, 4 Praesepe stars and 11 Hyades stars fall on the predicted failure modes that we corrected for following the method described in the last section. 
We then visually inspect all the available ZTF, TESS, and K2 light curves for each star with an accepted ZTF period. 
If the TESS or K2 period is deemed to be more reliable, we adopt that period instead of the ZTF measurement (9 in the Hyades). 
Stars with ambiguous period determinations are excluded from the sample (4 in the Hyades and 2 in the Praesepe). 
The final accepted sample consists of 26 stars in Praesepe and 100 stars in the Hyades.
Rotation period measurements that are $>35$ days, including the LS periodical and folded light curve from ZTF, TESS, and K2 when available are shown in \autoref{appx:longP}.

Finally, we include stars with literature-reported periods that also have ZTF detections, adopting the periods described in \autoref{subsec:protmeasure}. 
The resulting final sample consists of 123 stars in Praesepe and 121 stars in the Hyades.

\subsection{Photometric variability, binary identification, and \vsini\ measurements}\label{subsec:varibility}
For each star, we also measure the photometric variability, $R_{\rm var}$, by calculating the standard deviation of the median magnitude, $\tilde{m}$, over the entire light curve. 
We then convert the variability from magnitudes to relative flux units using $R_{\rm var} = 9.21 \times 10^5 \times {\rm std}(\tilde{m})\ {\rm ppm}$, where the constant factor converts variations in magnitude to relative flux variations under the assumption that the variability is small. 

To vet for potential binaries, we first selected stars with re-normalized unit weight error (RUWE) larger than the sky-varying RUWE threshold for binaries determined using the method described in \citep{CastroGindard2024}.
We also cross-matched our sample and the full cluster sample in \cite{Hunt2023} with SDSS-V IPL4 APOGEE observations \citep{Kollmeier2017, Wilson2019, Kollmeier2026} to obtain available radial velocity (RV) and \vsini\ measurements.
We identified confirmed binaries as stars with multiple good RV measurements from MWM and an RV scatter $>1$ km/s. 
Based on these criteria, we identify 6 confirmed binaries among the newly measured rotators in Praesepe and 1 in the Hyades.

\autoref{tab:1} shows the column description for our final catalog.
The period recovery fraction increases from 0\% at (\bprp)$\approx$1.5 to 100\% in Praesepe and 50\% in the Hyades at (\bprp)$\approx$3, before decreasing to 0\% at (\bprp)$\approx$3.25 in both clusters.
The difference in period recovery fraction for the Praesepe and Hyades at (\bprp)$\approx$3 could indicate larger field contaimination for the Hyades.
\begin{table*}[]
    \centering
    \caption{Column description for the full rotation period catalog for the Hyades and Praesepe. 
    This table is published in its entirety in a machine-readable format in the online journal.}
    \begin{tabular}{c|c}
         Column Name & Description  \\
         \hline\hline
         Name & Cluster Name (NGC 2632 for Praesepe and Melotte 25 for Hyades) \\
         Prot\_ztf & Measured ZTF \prot \\
         Prot\_K2 & Measured K2 \prot \\
         Prot\_TESS\_unpopular'& Measured TESS \prot\ using \texttt{unpopular} \\
         Prot\_TESS & TESS \prot\ from \cite{Boyle2026} \\
         Prot\_correct & ZTF \prot\ after correcting for failure modes (recommend using) \\
         m\_Prot\_k2\_correct & Whether the K2 \prot\ is used to correct for the ZTF failure modes \\
         m\_Prot\_tess\_correct & Whether the \cite{Boyle2026} TESS \prot\ is used to correct for the ZTF failure modes \\ 
         m\_Prot\_tess\_correct\_unpopular & Whether the \texttt{unpopular} TESS \prot\ is used to correct for the ZTF failure modes \\
         m\_Prot\_correct & Whether a ZTF failure mode is corrected \\
         GaiaDR3 & Gaia DR3 source ID \\
         ra & Gaia DR3 right ascension \\
         dec & Gaia DR3 declination \\
         pmRA & Gaia DR3 proper motion in the ra direction \\
         pmDE & Gaia DR3 proper motion in the dec direction \\
         phot\_g\_mean\_mag & Gaia DR3 $G$ magnitude \\
         phot\_rp\_mean\_mag & Gaia DR3 $G_{\rm RP}$ magnitude \\
         phot\_bp\_mean\_mag & Gaia DR3 $G_{\rm BP}$ magnitude \\
         parallax & Gaia DR3 parallax \\
         ruwe & Gaia DR3 re-normalized unit weight error \\
         binary & Gaia DR3 binary identification \\
         RV\_gaia & Gaia DR3 radial velocity \\
         sdss\_id & SDSS-V ID \\ 
         RV\_mwm & SDSS-V DR20 radial velocity \\
         v\_sini\_mwm & SDSS-V \vsini\ measurement \\
         P\_field & field contamination probability (\autoref{subsec:FieldCont}) \\
         2MASS & 2MASS ID \\
         mass\_mk & Mass estimated with \cite{Mann2019}
    \end{tabular}
    \label{tab:1}
\end{table*}

\section{Results} \label{sec:result}
In this section, we first validate the long rotation periods with $v \sin i$ measurements in \autoref{subsec:vsini}.
We then assess the potential impact of field-star contamination in \autoref{subsec:FieldCont}, specifically investigating whether the long rotation periods could be explained solely by contamination from non-member field stars. 
In \autoref{subsec:periodDist}, we examine the rotation-period distributions as a function of Gaia \bprp\ color, photometric temperature, and absolute magnitude using the newly measured ZTF periods.
We also investigate the activity–rotation relation in \autoref{subsec:activityRotation}.
Finally, in \autoref{sec:dynamo}, we discuss the implications of discovering these slowly rotating M dwarfs in open clusters younger than 1 Gyr. 
We explore the consequences for stellar dynamo theory and examine how the existence of such slow rotators may impact gyrochronology models and age determinations for FC stars.

\subsection{$v\sin i$ measurements of the slow-rotators}\label{subsec:vsini}
Because line broadening limits the precision of low \vsini\ measurements, values below $\sim10$ km s$^{-1}$ for slowly rotating stars are not expected to correlate strongly with rotation period \citep[e.g.,][]{Gilhool2018, Simonian2020}. 
We therefore compared our measured rotation periods with SDSS-V \vsini\ measurements to verify that the candidate slow rotators do not show anomalously large projected rotational velocities.
However, it is worth noting that most of the stars with newly measured rotation periods in our sample are K and M dwarfs, for which \vsini\ measurements may be artificially inflated to compensate for inadequately modeled spectral lines. 
We therefore regard the \vsini\ measurements as supporting evidence for the slow rotation of these stars, while noting that they should not be treated as an independent, quantitative constraint on their rotation periods.

\begin{figure}[ht!]
\centering
\includegraphics[width=\columnwidth]{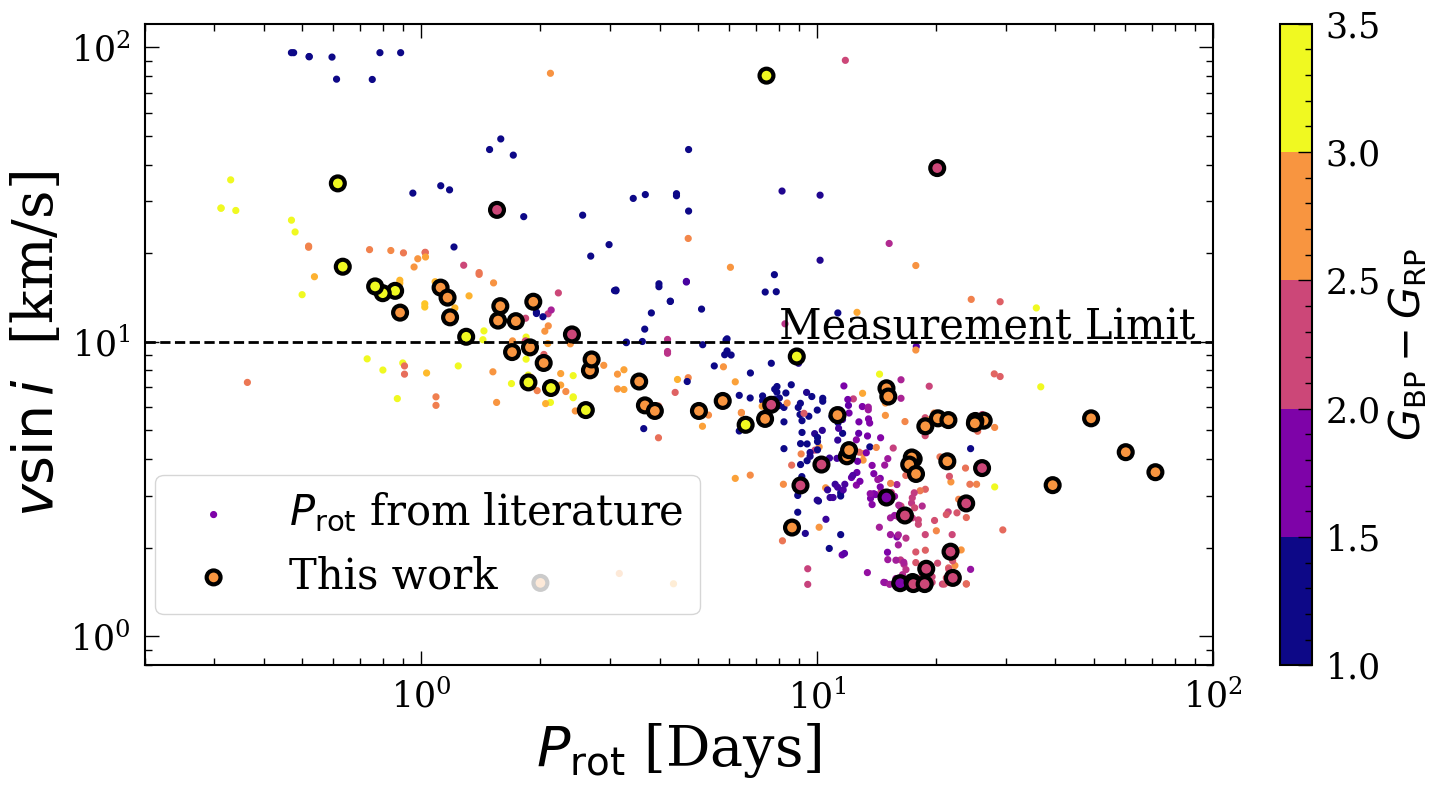}
\caption{Comparison of $v\sin i$ and rotation period for stars in Praesepe and Hyades. 
Stars with rotation periods from the literature are shown as smaller dots, while stars with newly measured rotation periods are shown as circles with black outlines. 
Points are colored by their \bprp\ color. 
Higher-mass hotter stars (smaller \bprp) exhibit a tighter $v\sin i$--$P_{\rm rot}$ relation, while lower-mass stars show larger scatter, likely due to systematic inflation of $v\sin i$ measurements from inadequately modeled spectral lines. 
Our newly measured rotation periods follow the same general trend as literature measurements for stars of similar \bprp. 
Stars with $P_{\rm rot}>35$ days have low $v\sin i$ values $<$ 10 km/s, consistent with the expected low projected rotational velocities of slowly rotating stars and the apparent $v\sin i$ detection floor.
\label{fig:6}}
\end{figure}

\autoref{fig:6} shows the comparison for both clusters, with stars with literature rotation periods shown as crosses and stars with newly measured rotation periods shown as circles. 
All stars are colored by their \bprp\ color. 
Higher-mass stars (smaller \bprp) exhibit a stronger and tighter \vsini-\prot\ relation than lower-mass stars (larger \bprp), as expected, likely because \vsini\ measurements for lower-mass stars can be artificially inflated to compensate for inadequately modeled spectral lines. 
Our newly measured rotation periods follow a similar declining trend in \vsini\ with increasing \prot, and are consistent with stars of similar \bprp\ color with rotation periods from the literature.
Stars with rotation periods $>35$ days appear to deviate from the trend, likely because the literature sample contains few stars in this period regime. 
Nevertheless, all of these stars have \vsini\ measurements $<10$ km s$^{-1}$, consistent with the expected low projected rotational velocities of slowly rotating stars and the measurement limit of \vsini. 
This floor is particularly evident for stars with \prot $> 10$ days and \bprp\ $>2.5$, for which \vsini\ measurements cluster near the detection limit.

\subsection{Field contamination}\label{subsec:FieldCont}
One important question is whether these slowly rotating stars could instead be field-star contaminants. 
In this section, we demonstrate that this concern is valid for some of our longest rotation period stars. 
However, we do find a significant sample of stars where the background hypothesis is extraordinarily unlikely.
To investigate this, we first use the 6D kinematic information from Gaia DR3 and SDSS-V. 
None of the slow rotators show significant deviations in their 5D kinematics (excluding RV) from the rest of the cluster sample, as expected given that they were identified using 5D kinematic membership information.
For the stars with RV measurements, we likewise find no significant kinematic deviations between the slow rotators and the rest of the Hyades sample.
We do not have RV measurements for the slow rotators in Praesepe as these stars are too faint for reliable spectroscopic measurements.

\begin{figure*}
    \centering
    \includegraphics[width=\textwidth]{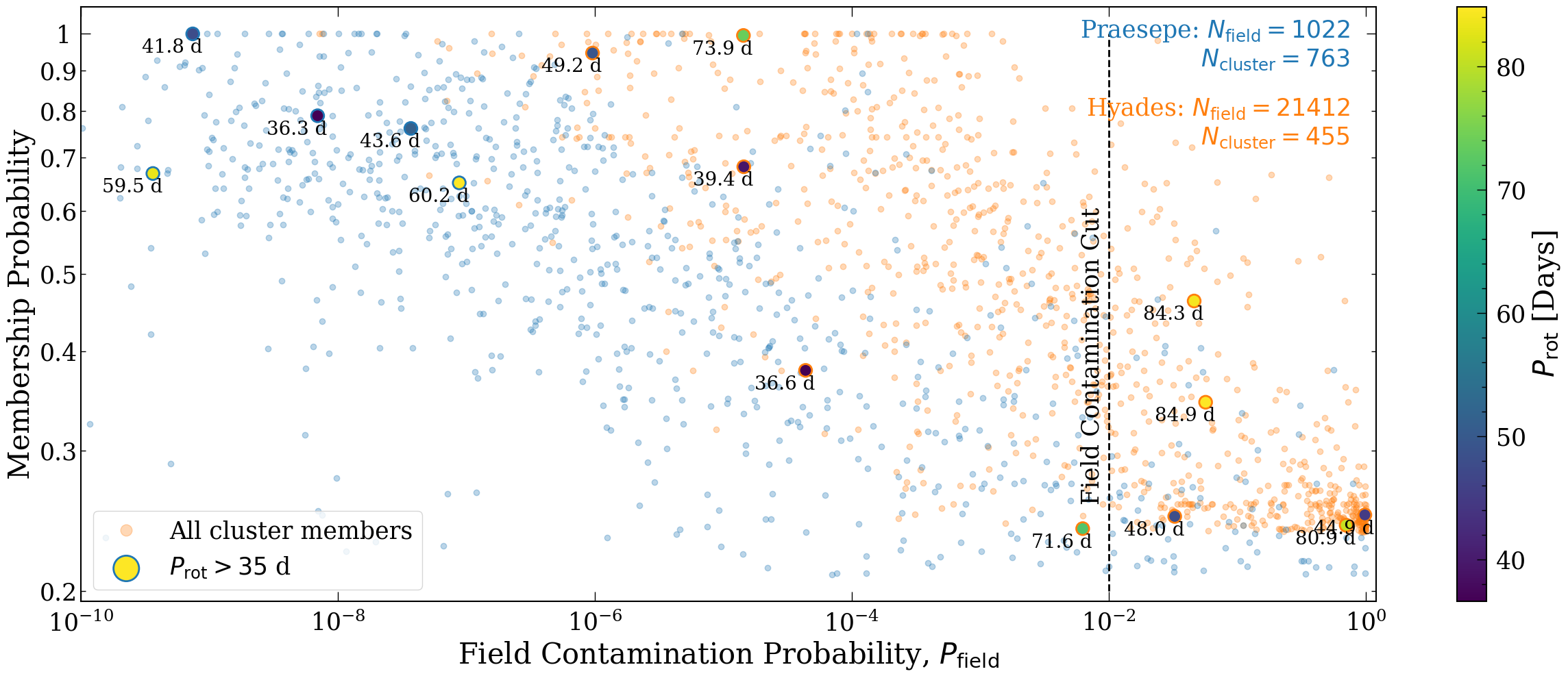}
    \caption{Field contamination probability ($P_{\rm field}$) for the cluster members.
    The probability that each star is a field contaminant, $P_{\rm field}$, is estimated using Gaussian mixture models of the cluster and field populations in astrometric and photometric parameter space. 
    Stars with available radial velocities are evaluated using 6D kinematics together with $G$ magnitude and \bprp, while stars without radial velocities are evaluated using 5D astrometry and the same photometric quantities. 
    The resulting likelihoods are combined with the expected cluster and field fractions of the candidate sample using \autoref{eq:2}. 
    Hyades and Praesepe members are shown in orange and blue, respectively, with the long-period stars ($P_{\rm rot}>35$ days) highlighted by larger symbols and colored according to their rotation periods.
    The number of expected total number of field and cluster stars with similar 5D kinematics used to calculate the priors are shown on the top right corner.
    5/10 stars in the Hyades and 0/5 stars in the Praesepe have $P_{\rm field}>0.01$.
\label{fig:distance_cont}}
\end{figure*}

We further estimated the probability of field contamination, $P_{\rm field}$, for each cluster member using Gaussian mixture models to separately model the cluster and field populations in astrometric and photometric parameter space. 
For each star with observed properties $\mathbf{x}$ (in this case, 5D or 6D kinematics and Gaia $G$ and \bprp), the probability of being a field star is given by Bayes' theorem, 
\begin{equation}\label{eq:2}
    P_{\rm field}(\mathbf{x}) =\frac{p(\mathbf{x}\mid{\rm field})P({\rm field})}{p(\mathbf{x}\mid{\rm field})P({\rm field})+p(\mathbf{x}\mid{\rm cluster})P({\rm cluster})},
\end{equation}
where $p(\mathbf{x}\mid{\rm field})$ and $p(\mathbf{x}\mid{\rm cluster})$ are the likelihoods of the star's observed properties under the field and cluster Gaussian mixture models, respectively, and $P({\rm field})$ and $P({\rm cluster})$ are the adopted prior probabilities.
We estimate the prior cluster and field fractions from the membership probabilities of the candidate sample and the total number of field sample with similar 5D kinematics ($N_{\rm f,5D}$). 
Specifically, the sum of the membership probabilities, $\sum_i P_{{\rm mem},i}$, gives the expected number of cluster members ($N_{\rm cluster}$), while $\sum_i(1-P_{{\rm mem},i})$ gives the expected number of field contaminants in the cluster sample. 
$N_{\rm field}$ is then $\sum_i(1-P_{{\rm mem},i})+N_{\rm f,5D}$.
$P({\rm field})$ and $P({\rm cluster})$ are then calculated as $N_{\rm field}/(N_{\rm field}+N_{\rm cluster})$ and $N_{\rm cluster}/(N_{\rm field}+N_{\rm cluster})$, respectively.
We use the corresponding fractions of the total sample as the priors in the Bayesian contamination calculation.
We construct the control-field sample by selecting stars within the smallest region in RA, Dec, proper motion, and parallax space that encompasses the cluster members.
For stars with available radial velocities, the models incorporate the full 6D kinematic information together with $G$ magnitude and \bprp; for stars without radial velocities, we use the corresponding 5D astrometric information. 
The resulting contamination probabilities are shown in \autoref{fig:distance_cont}. 
Including radial velocities where available does not substantially change the contamination estimates or the resulting interpretation of the long-period stars. 
Five of the ten long-period Hyades stars have relatively high $P_{\rm field}$ ($>$0.01), indicating that they could be field contamination. 
This removes all stars with \prot$>$80 days.
In contrast, the five long-period Praesepe stars and rest of the five Hyades stars have low $P_{\rm field}$, providing stronger evidence that they are genuine cluster members. 
Notably, several of these likely cluster members have rotation periods of $\sim60$--70 days.

We also cross-matched the field sample with similar 5D kinematics to the clusters against the field rotation-period catalog described in \autoref{sec:data_method}. 
We find that 17\% and 3\% of the field stars around the Hyades and Praesepe, respectively, have detected rotation periods.
Using these samples, we performed a Monte Carlo analysis to quantify the expected contribution of field contamination to the observed slow-rotator population. 
We first constructed mock cluster CMDs from the corresponding field samples by selecting, for each cluster star, a random star from the 50 closest field stars in \bprp\ and Gaia $G$ magnitude. 
We then calculated the fraction of slow rotators ($P_{\rm rot}>35$ days) in each mock cluster sample, noting that most field stars do not have measured rotation periods.
We repeated this procedure 5000 times to obtain the distribution of slow-rotator fractions expected from field contamination. 
Finally, we compared this distribution with the membership-probability-weighted slow-rotator fraction measured in the actual cluster sample.
None of the 5000 mock Praesepe samples contains a slow rotator, indicating that the observed Praesepe slow-rotator population cannot be reproduced by the field sample in our Monte Carlo realizations. 
For the Hyades, the observed slow-rotator fraction lies $5.6\sigma$ above the distribution expected from field contamination.
This provides strong evidence that the observed slow-rotator population is an intrinsic property of the clusters rather than a result of field-star contamination.

We also examine the period distribution in different $M_G$ bins for different membership probability thresholds, assessing whether it provides a reliable criterion for identifying cluster members. 
We then compare the fraction of stars with  \prot$> 35$ days as a function of \bprp\ between the field and cluster samples to determine whether the field population alone can explain the observed population of slow rotators.

\begin{figure}[ht!]
\centering
\includegraphics[width=\columnwidth]{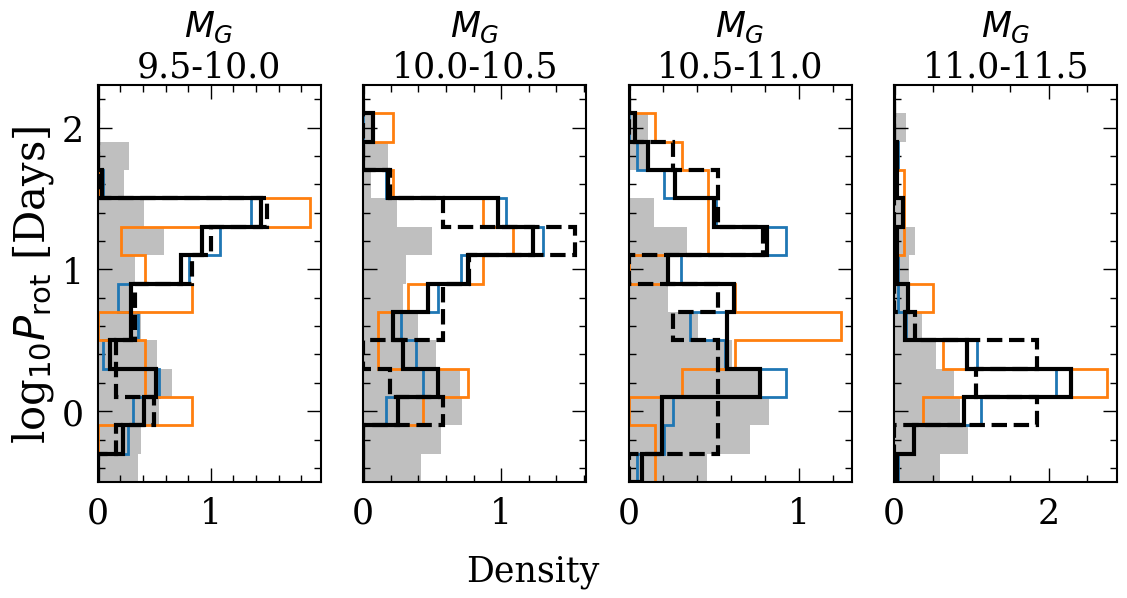}
\includegraphics[width=\columnwidth]{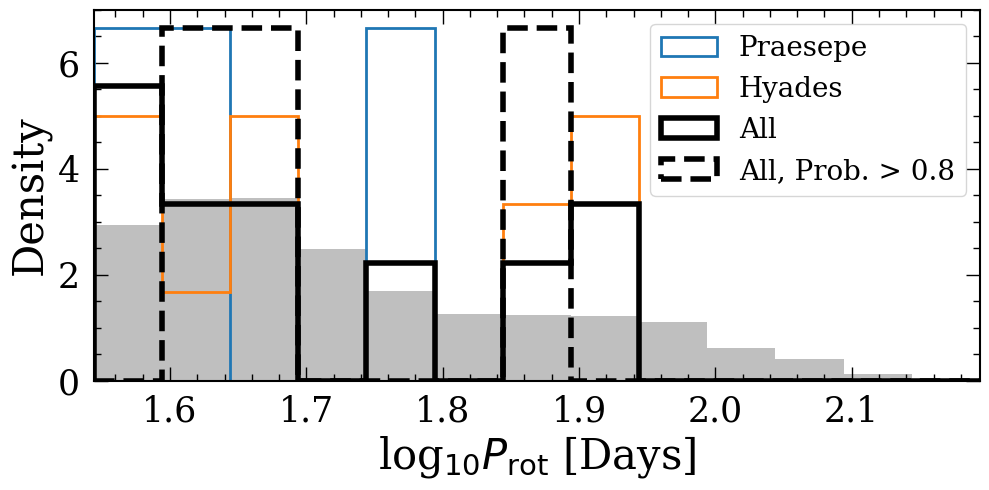}
\caption{Top: Normalized density of rotation period, \prot, as a function of absolute magnitude, $M_G$, in 0.5 mag bins. 
Field stars are shown in gray, while the full Praesepe and Hyades samples, including periods newly identified in this work and periods from the literature \citep{vanlane2025}, are shown in orange and blue, respectively. 
The combined cluster sample is shown by the solid black line, while the sample restricted to stars with membership probability $>0.8$ is shown by the dashed black line. 
The cluster rotation-period distributions are similar with and without the membership-probability cut, indicating that the observed distributions are insensitive to this additional selection criterion.
In particular, the slowly rotating population persists with little change in the $M_G=10.5$–11 bin regardless of the membership-probability cut, suggesting that these stars are unlikely to be primarily field-star contaminants. 
Bottom: Rotation-period distributions for stars with \prot$>35$ days in the cluster and field samples. 
The cluster distribution does not closely match that of the field population, providing further evidence that the slow-rotating M dwarfs are not simply field-star contaminants. 
We caution, however, that the sample sizes are small, with only 18 and 3 stars having \prot$>35$ days before and after applying the membership-probability cut, respectively, limiting our ability to robustly characterize the distribution of this population.
\label{fig:10}}
\end{figure}

\autoref{fig:10} top panel shows the normalized density in \prot\ as a function of $M_G$ in bins of 0.5 mag. 
Field stars are shown in gray, while the full Praesepe and Hyades samples, including periods newly identified in this work and periods from the literature, are shown in orange and blue, respectively, without applying a membership probability cut. 
The combined cluster sample is shown by the solid black line, while the sample restricted to stars with membership probability $>0.8$ is shown by the dashed black line.
It is clear that the cluster distributions as a function of $M_G$ are very similar with and without the membership probability cut. 
This suggests either that membership probability is not an effective criterion for vetting cluster members in these two clusters, or that the presence of a measurable rotation period itself provides a stronger membership constraint, such that stars with reliable rotation period measurements are already likely to be cluster members.
In addition, the slowly rotating stars persist with little change in their distribution in the $M_G=10.5$–11 bin, regardless of the membership-probability cut. 
This suggests that their presence is likely real and is not primarily driven by field-star contamination. 
Finally, the period distribution of stars with $P_{\rm rot}>35$ days in the clusters (bottom panel of \autoref{fig:10}) does not closely match that of the field population, further suggesting that the slow-rotating M dwarfs are not simply field-star contaminants. 
However, we caution that the sample sizes are small, with only 18 and 3 stars having $P_{\rm rot}>35$ days before and after applying the membership-probability cut, respectively, making it difficult to robustly characterize the period distribution of this population.

\begin{figure}[ht!]
\centering
\includegraphics[width=\columnwidth]{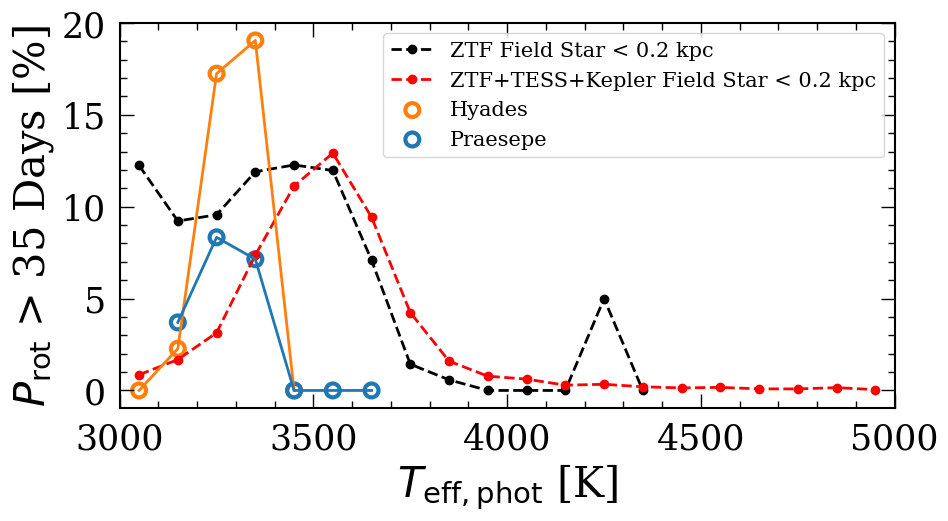}
\caption{Fraction of stars with \prot$>35$ days as a function of effective temperature for the ZTF field sample, the full field sample combining rotation periods from all surveys as descirbed in \autoref{subsec:opencluster}, and the Hyades and Praesepe cluster samples. 
The field samples are shown both with a 200 pc distance cut to better match the cluster sample. 
All four samples exhibit a peak in the fraction of slow rotators, with the field samples peaking at $\sim3500$ K and the Hyades and Praesepe peaking at $\sim3300$ K. 
The different peak locations not only suggest that the slow rotators in the clusters are unlikely to be drawn from the field population, but may also reflect the metallicity dependence of the fully convective boundary, which is expected to occur at lower temperatures for more metal-rich stars.
The broader peak in the field samples may result from their wider metallicity distribution, while the narrower peaks in the clusters may reflect their more uniform metallicities. \label{fig:11}}
\end{figure}

In addition, we compare the fraction of long-period rotators as a function of photometric temperature between field stars and the two clusters. 
We calculate the fraction of stars with \prot$>35$ days relative to the total number of stars in 100 K wide temperature bins, using the ZTF-only field sample and the full field sample combining rotation periods from all surveys. 
For both field samples, we restricted to within 200 pc to better match with the cluster sample.
We use the ZTF-only sample as a comparison because most of the long-period rotation measurements in the cluster samples come from ZTF. 
We perform the same calculation separately for the Hyades and Praesepe clusters.
\autoref{fig:11} shows the results. 
All four samples exhibit a clear peak in the fraction of stars with \prot$>35$ days.
The location of this peak does not change significantly when adopting \prot$>$40 or 50 days, or when not imposing a distance cut in the field sample, or if restricting the field samples to within 100 pc.
Interestingly, the peak occurs at different temperatures for the field and cluster samples.
Both the ZTF-only and full field samples peak at around 3500 K, while the Hyades and Praesepe samples peak at a slightly lower temperature, around 3300 K. 
This difference further suggests that the slowly rotating stars in the clusters are not simply drawn from the field population, as field contamination would be expected to produce a similar temperature dependence.

As a thought experiment, if the peak is associated with the FC boundary, this raises the possibility that the location and width of the slow-rotator peak could provide a new way to probe the metallicity distribution of a stellar population. 
The FC boundary is predicted to occur at $\mathrm{[Fe/H]}\sim-0.1$ at $\sim3500$ K and $\mathrm{[Fe/H]}\sim+0.3$ at $\sim3300$ K \citep[see Figure 1 in][]{Lu2024}, consistent with the more metal-rich nature of the Hyades and Praesepe compared to the field population.
Since the field sample spans a broader range of metallicities, it is also plausible that its slow-rotator peak is broader than those of the clusters. 
In addition, a transition in the temperature–radius relation has been observed around 3200–3300 K \citep{Rabus2019}, corresponding to a mass of approximately $0.2,M_\odot$, which may also be associated with the FC boundary and coincides with the location of the peaks in the Hyades and Praesepe.

\subsection{Period Distribution}\label{subsec:periodDist}

After establishing that field-star contamination is unlikely to fully explain the long-period population, in this section, we first examine the rotation periods in the context of the CMD, and then investigate the rotation-period distribution as a function of color and effective temperature by combining our measurements with those from the literature.

\begin{figure*}[ht!]
\centering
\includegraphics[width=\textwidth]{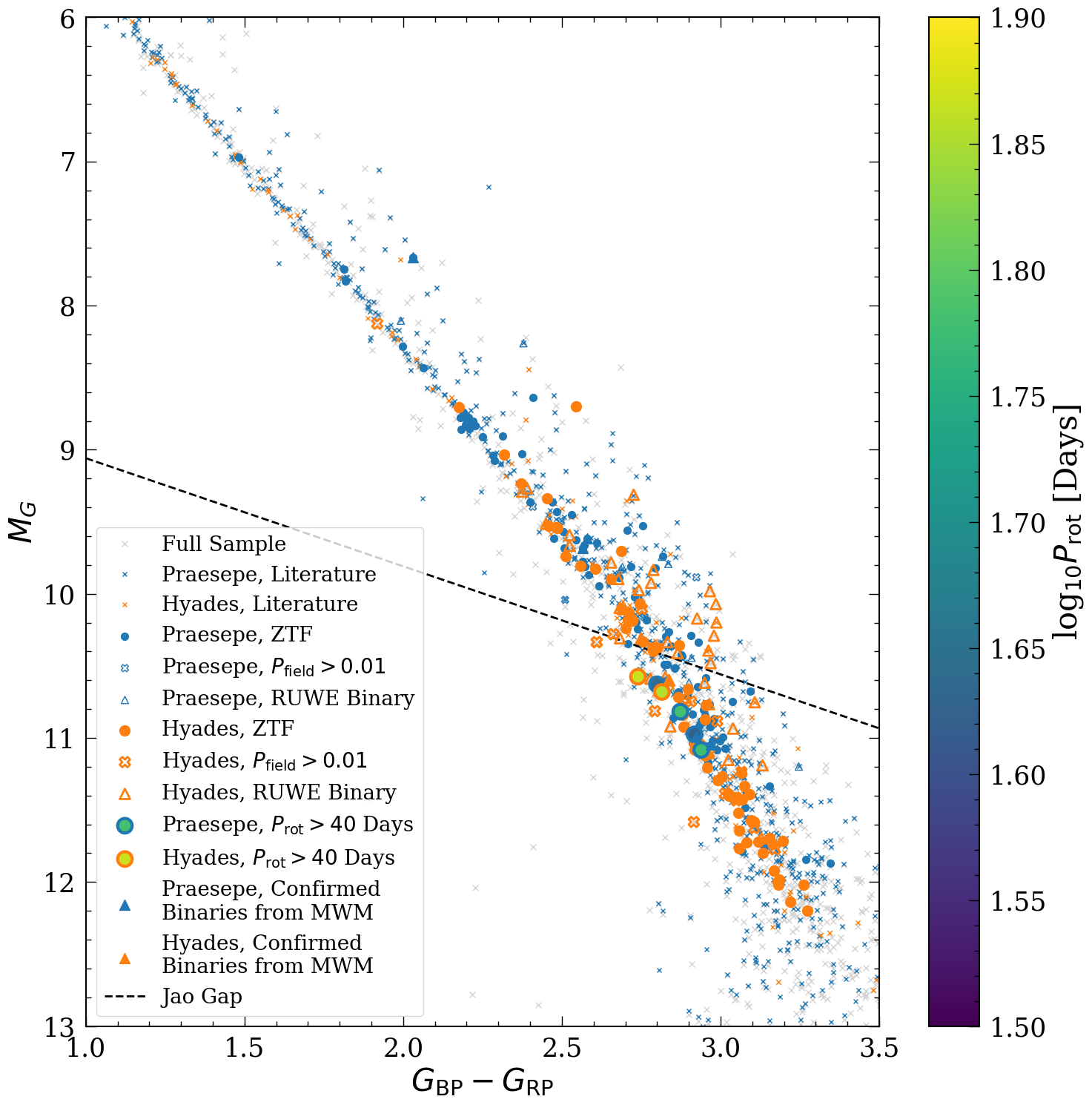}
\caption{Color-magnitude diagram of the Praesepe and Hyades cluster stars searched for rotation periods. 
Gray crosses show the full sample searched for rotation periods, as in the left panel of \autoref{fig:1}.
Blue and orange crosses denote stars with rotation periods from the literature, while circles and triangles denote stars with rotation periods measured in this work that are classified as single and potential binaries, respectively, based on their Gaia RUWE values. 
The empty crosses show stars with field contamination probability ($P_{\rm field} >0.01$).
Stars with measured rotation periods $>40$ days are colored by their rotation periods. 
The black dashed line indicate the Jao gap, separating the FC and partially convective (PC) stars in the CMD.
Stars with measured rotation periods form a tight sequence in the CMD, consistent with a coeval stellar population. 
Potential binaries are predominantly overluminous relative to the main sequence, as expected for unresolved multiple systems. 
The slowly rotating stars preferentially lie toward the hotter edge of the sequence, consistent with weaker magnetic activity due to their slow rotation and thus, reduced radius inflation.
It is also interesting that the most slowly rotating stars preferentially lie near the Jao gap.
\label{fig:7}}
\end{figure*}

\autoref{fig:7} shows the CMD for the cluster stars, where the gray crosses show the full sample we searched for rotation periods for (same as left panel of \autoref{fig:1}), the blue and orange crosses show those with literature periods, and the circles and tangles show the single and binary determined with RUWE, respectively, for stars with period measurements from this work. 
The empty crosses show stars with field contamination probability ($P_{\rm field} >0.01$).
Stars with measured period $>$ 40 days are colored by their period measurements. 
It is clear that the stars with measured rotation periods lie along a tight sequence, consistent with expectations for a coeval stellar population. 
In addition, stars identified as potential binaries are predominantly overluminous relative to the main sequence, as expected for unresolved multiple systems. 
The slowly rotating stars preferentially lie toward the hotter edge of the sequence, consistent with their weaker magnetic activity associated with their rotation and, consequently, reduced radius inflation.
The Jao gap \citep{Jao2018}, defined by the line connecting ($M_G$, \bprp) $\sim$ (10.09, 2.375) and ($M_G$, \bprp) $\sim$ (10.24, 2.575), is indicated by the black dashed line. 
The gap corresponds to an under-density of stars in the CMD and is thought to arise from a $^3$He-driven structural instability as stars transition from partially to fully convective \citep{vansaders2012}.
It is interesting that the most slowly rotating stars preferentially lie near the Jao gap. 
This could indicate that the transition from partially to fully convective interiors is associated with a change in the efficiency of magnetic braking, potentially allowing stars near this transition to reach unusually long rotation periods \citep[e.g.,][]{Chiti2024}. 

\begin{figure*}[ht!]
\centering
\includegraphics[width=0.99\textwidth]{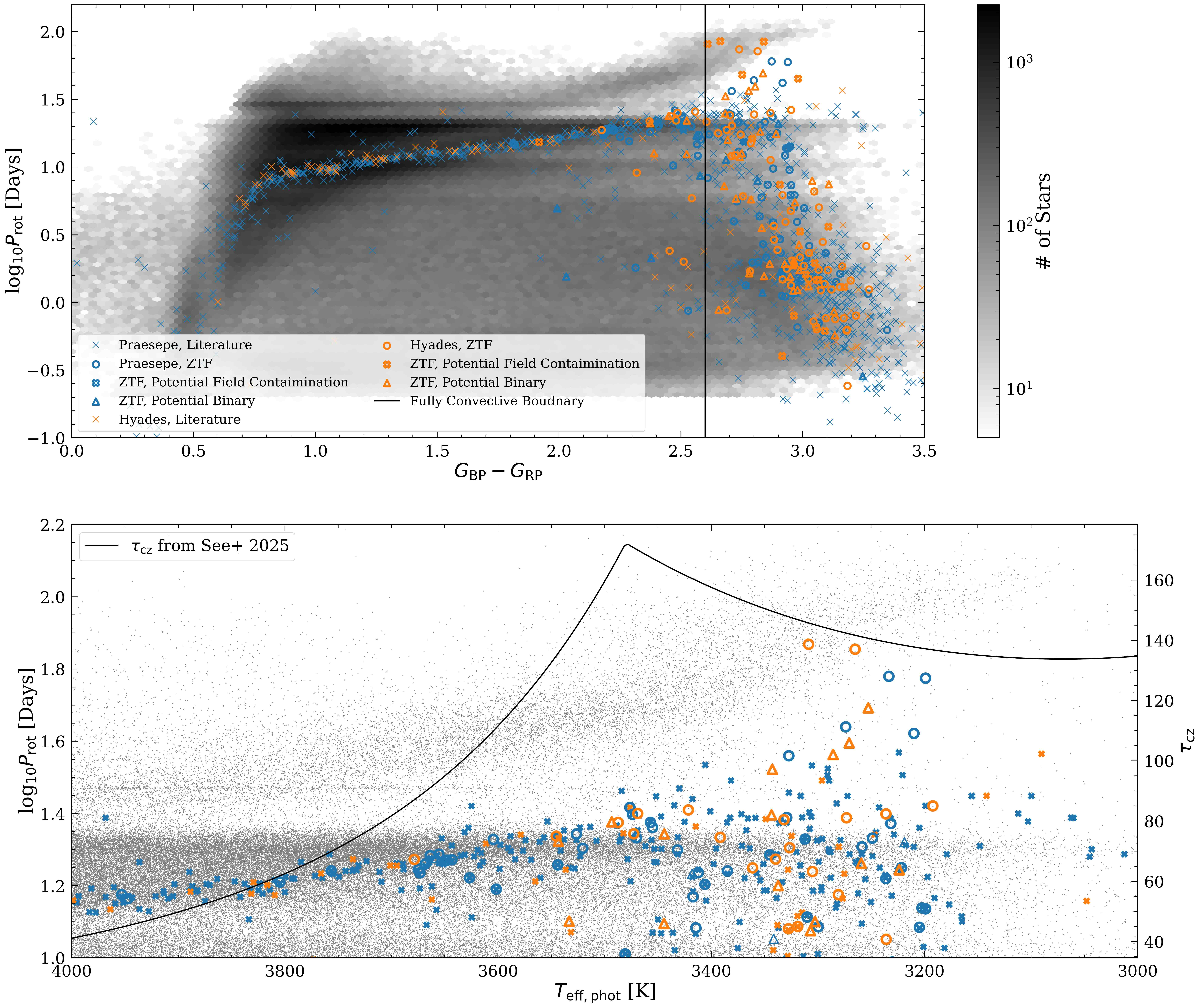}
\caption{Rotation period distribution of Praesepe and Hyades stars, combining measurements from this work and the literature. 
Crosses denote literature rotation periods, while circles and triangles denote periods measured in this work for stars classified as single and potential unresolved binaries, respectively, based on their Gaia RUWE values. 
The empty crosses show stars with field contamination probability ($P_{\rm field} >0.01$).
The background shows the period distribution of field stars. 
The abrupt cutoff at $P_{\rm rot}=15$ days reflects the $P_{\rm rot}<15$ day selection applied to the TESS field-star sample, while the feature near $P_{\rm rot}\sim29.5$ days corresponds to a ZTF systematic associated with the lunar cycle. 
The absence of stars with \bprp\ $<1.5$ from ZTF is primarily due to saturation.
The bottom panel shows a zoomed-in view of the slowly rotating M dwarfs, with the $x$-axis showing photometric effective temperature inferred from \bprp\ using the calibration of \cite{Curtis2020}. 
Newly measured rotation periods are indicated by open markers. 
The triangles show potential binaries.
The background again shows the field-star rotation-period distribution, and the solid line shows the convective turnover time, \tauc, as a function of effective temperature from \cite{See2025}. 
The vertical transition near $\sim3500$ K marks the approximate fully convective boundary, where the field-star rotation-period distribution becomes more diffuse and extends to longer periods. 
The slowly rotating stars identified in Praesepe and Hyades occupy the same region of period--temperature space as the slowly rotating fully convective field-star population. 
\label{fig:8}}
\end{figure*}

\autoref{fig:8} shows the period-temperature distribution, with crosses denoting literature measurements and circles and triangles denoting periods measured in this work.
Triangles indicate stars that may be unresolved binaries based on their Gaia RUWE values and empty crosses show stars with field contamination probability ($P_{\rm field} >0.01$).
For comparison, we also show the rotation-period distribution of field stars in the background. 
The abrupt cutoff in the field-star distribution at \prot\ = 15 days results from our restriction to \prot\ $<$ 15 days for the TESS sample, as periods longer than 15 days require additional visual vetting to ensure their reliability \citep{Boyle2026}. 
The horizontal feature near \prot$\sim29.5$ days corresponds to a systematic in the ZTF data associated with the lunar cycle.
Our sample does not include stars with \bprp\ $<$ 1.5, as most higher-mass stars are saturated in ZTF. 

At the low-mass end, the additional sample of rapidly rotating stars identified from ZTF is consistent with the literature measurements, further demonstrating the reliability of our rotation period measurements. 
The difference becomes apparent at \prot\ $>$ 35 days, where no stars in the literature occupy this period range. 
This is most likely due to the short observing season of $\sim$70 days for K2, which limits its sensitivity to long rotation periods, as demonstrated in \autoref{subsec:protmeasure}.
Interestingly, long rotation periods are detected only for stars around or beyond the FC boundary \citep[\bprp\ $\sim$ 2.6;][]{Lu2022}. 

The bottom panel of \autoref{fig:8} shows a zoomed-in view of the slowly rotating M dwarfs from the top panel. 
The $x$-axis shows the photometric effective temperature inferred from \bprp\ using the coefficients from \cite{Curtis2020}.
As in the top panel, crosses denote rotation periods from the literature, while circles and triangles denote periods measured in this work for stars classified as single and potential binaries, respectively, based on their Gaia RUWE values. 
The newly measured periods are colored by their cluster membership probability. 
The filled crosses show stars with field contamination probability ($P_{\rm field} >0.01$).
The background shows the rotation-period distribution of field stars for comparison. 
We also show the convective turnover time (\tauc) as a function of effective temperature, adopted from \cite{See2025}.
The convective turnover time peaks at $\sim3500$ K, corresponding approximately to the FC boundary, where stars transition to fully convective interiors and the convective turnover time increases rapidly toward lower temperatures.
Passed the FC boundary, \tauc\ decreases due to the decrease in radius.
The FC boundary also coincides with a transition in the rotation period distribution of field stars. 
PC stars exhibit a clear and relatively tight sequence, with $\log_{10}$\prot\ increasing to $\sim1.5$, whereas the period distribution of FC stars is more diffuse and extends beyond $\log_{10}$$ P_{\rm rot}>1.5$ \citep[see further discussion in][]{Lu2024}.
The slow-rotating stars in Praesepe and Hyades all fall within the rotation period distribution of FC field stars.

Many of the long-period rotators have low membership probabilities, however, as discussed in \autoref{subsec:FieldCont}, these low probabilities may not reliably indicate cluster membership and could instead be driven by uncertainties in their photometric and astrometric measurements.
Also as shown in \autoref{fig:1}, the average membership probability decreases toward lower-mass stars in the CMD. 
A low membership probability therefore does not necessarily indicate that these stars are non-members, but rather reflects the increased difficulty of distinguishing faint cluster members from field stars.
Nevertheless, even when restricting the sample to stars with membership probabilities $>0.5$, several slowly rotating, fully convective M dwarfs remain. 
If these stars are confirmed cluster members, their presence would suggest that fully convective stars can undergo extremely rapid magnetic braking, reaching very long rotation periods within $<1$ Gyr.

\subsection{The activity-rotation relation}\label{subsec:activityRotation}
Stellar activity is known to be correlated with a star's rotation period through the \ro--activity relation.
The behavior of the \ro--activity relation across the fully convective boundary is therefore particularly relevant for understanding the long rotation periods identified in our sample.
If the slowly rotating fully convective stars have unusually large Rossby numbers, their weak magnetic activity could indicate a reduced efficiency of angular-momentum loss. 
Conversely, if these stars remain magnetically active despite their long rotation periods, this would suggest that the standard \ro--activity relation does not adequately describe the magnetic behavior of fully convective stars. 
We therefore compare the activity of our cluster stars with their rotation periods to investigate whether the slowly rotating stars exhibit anomalous magnetic activity near and beyond the fully convective boundary.

\begin{figure*}[ht!]
\centering
\includegraphics[width=\textwidth]{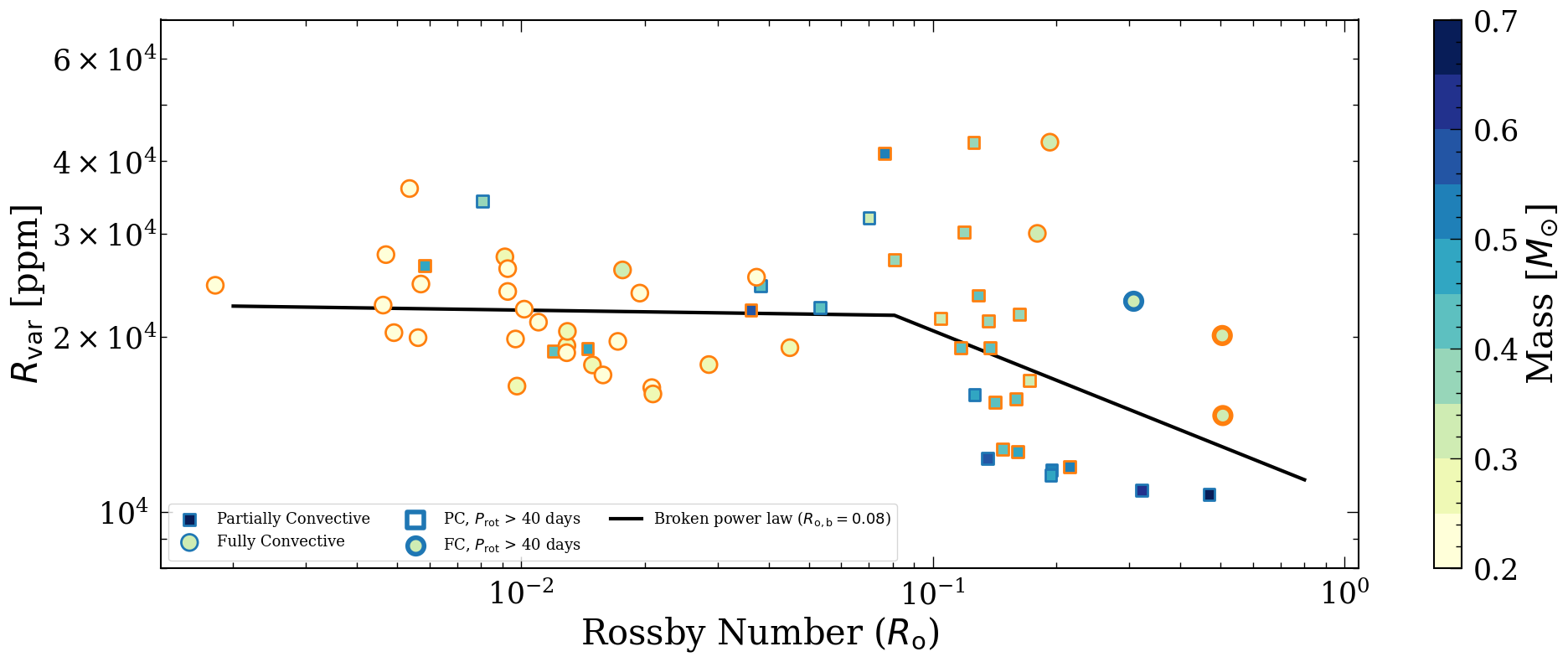}
\caption{The \rvar--\ro\ relation, where \rvar\ is measured as described in \autoref{subsec:protmeasure} and \ro\ is calculated from \prot\ using \tauc\ from \autoref{fig:8}. 
Blue and orange outlines denote Praesepe and Hyades stars, respectively. 
The squares and circles denote FC and PC stars, respectively, classified using the Jao gap (see \autoref{fig:7}). 
Stars with $P_{\rm field} > 0.01$, binaries based on RUWE and SDSS-V RVs, and $G > 16.5$ mag are excluded.
Marker colors indicate stellar masses estimated with a luminosity--mass relation \citep{Mann2019}. 
Bold outlines indicate stars with \prot$>40$ days.
The black line shows a broken power-law fit: $R_{\rm var}=2.18\times10^4(R_{\rm o}/0.081)^{-0.01}$ for $R_{\rm o}<0.081$ and $R_{\rm var}=2.18\times10^4(R_{\rm o}/0.081)^{-0.28}$ for \ro$>0.081$.
Despite the small sample size and large intrinsic scatter \citep[e.g.,][]{Santos2019}, an activity plateau is apparent for $\ro\lesssim0.1$, followed by a decline in \rvar\ toward larger \ro. 
FC stars appear to have higher \rvar\ than PC stars in the unsaturated regime, consistent with recent ZDI results showing enhanced large-scale magnetic fields in slowly rotating FC stars \citep{See2025}. 
This differs from the similar X-ray and H$\alpha$--\ro\ relations reported for FC and PC field stars \citep[e.g.,][]{Wright2016,Kiman2021,Nunez2022}.
However, previous cluster studies largely probe the saturated regime, while field-star FC samples are small and exhibit substantial scatter \citep[e.g.,][]{West2015,Newton2017}. 
Given similar limitations both from small sample size and the large intrinsic scatter in \rvar, we refrain from drawing strong conclusions about the FC/PC difference.
\label{fig:9}}
\end{figure*}

\autoref{fig:9} plots the \ro--$R_{\rm var}$ relation, with $R_{\rm var}$ measured in \autoref{subsec:varibility}, and \ro\ converted from \prot\ using \tauc\ plotted in \autoref{fig:8}.
The blue outlines represents stars from the Praesepe, and the orange represents those from the Hyades.
We excluded stars with field contamination $>$ 0.01 and binaries identified from RUWE and SDSS-V RV mesurements.
We also excluded stars with Gaia $G$ fainter than 17 mag as the photometric variability started to get dominated by noise.
The squares and circles represent FC and PC stars determined using the Jao gap (reference to \autoref{fig:7}), respectively.
The markers are colored by stellar mass, estimated using the mass--luminosity relation. 
We first cross-matched our sample with 2MASS and calculated the absolute $K$-band magnitudes, which we then used to estimate stellar masses using the fifth-order polynomial fit presented in \citet{Mann2019}.
Stars with bold outlines indicate the ones with rotation $>$ 40 days. 
Note that most of our stars with measured periods are fully convective ($M < 0.35$\mdot).
The black line indicates a broken power-law fit to the sample. 
We use \texttt{scipy.optimize} to minimize $\chi^2$, simultaneously fitting the break in \ro, the amplitude, and the slopes of the two power-laws.
The final fit is $R_{\rm var}=2.18\times10^4(R_{\rm o}/0.081)^{-0.01}$ for $R_{\rm o}<0.081$ and $R_{\rm var}=2.18\times10^4(R_{\rm o}/0.081)^{-0.28}$ for \ro$>0.081$. 

The sample size is small, and the scatter in the \rvar--\ro\ relation is known to be large \citep[e.g.,][]{Santos2019}; nevertheless, an activity plateau is still apparent for \ro$\lesssim0.1$, indicating the saturated regime.
At \ro$\gtrsim0.1$, \rvar\ decreases with increasing \ro, consistent with the unsaturated regime. 
In the unsaturated regime, FC stars appear to lie systematically above PC stars, consistent with the recent 15 pc M-dwarf study showing that the photometric amplitudes of fully convective M dwarfs do not depend strongly on rotation \citep{Pass2023}. 
This is also consistent with studies suggesting differences in magnetic-field topology and angular-momentum loss between FC and PC stars \citep[e.g.,][]{Lu2024,See2025}.
However, this appears to be in tension with previous studies of magnetic activity indicators, such as X-ray and H$\alpha$ emission, which generally find similar activity--\ro\ relations for PC and FC stars \citep[e.g.,][]{Wright2016,Kiman2021,Nunez2022}. 
It is worth noting that previous studies of magnetic activity in the Praesepe and Hyades have been largely limited to the saturated regime as the longest measured rotation periods are $\sim30$ days, corresponding to $\ro\sim0.2$ for a 3500 K star. 
Thus, the absence of a clear difference between FC and PC stars in these studies does not necessarily rule out a difference in the unsaturated regime, as the FC stars in these young clusters provide limited leverage on the activity--\ro\ relation beyond the saturation threshold. 
Given the large intrinsic scatter in the activity--\ro\ relation and the uncertain transition between the saturated and unsaturated regimes at $\ro\sim0.1$, the behavior of FC stars in the unsaturated regime remains difficult to constrain. 
Furthermore, field-star samples of FC stars are typically small and exhibit substantial non-detections or scatter at large \ro\ \citep[e.g.,][]{West2015, Newton2017}. 
However, our sample is similarly limited in size and exhibits considerable scatter, and we therefore refrain from drawing strong conclusions about the difference between FC and PC stars in the unsaturated regime.

More generally, the choice of rotational variable may itself affect the apparent form of activity--rotation relations. 
\citet{Stassun2024} found that the empirical relation between X-ray activity and rotation in young stars becomes substantially simpler when rotation is expressed in terms of stellar angular momentum rather than $P_{\rm rot}$ alone. 
This raises the possibility that some of the apparent mass dependence, and perhaps some of the structure across
the FC boundary, in period- or Rossby-based activity relations reflects the strong mass dependence of stellar structure. 
A future comparison of the present activity measurements directly against stellar angular momentum could therefore provide a complementary test of whether FC and PC stars obey intrinsically different activity--rotation relations.

\begin{figure*}[ht!]
\centering
\includegraphics[width=\textwidth]{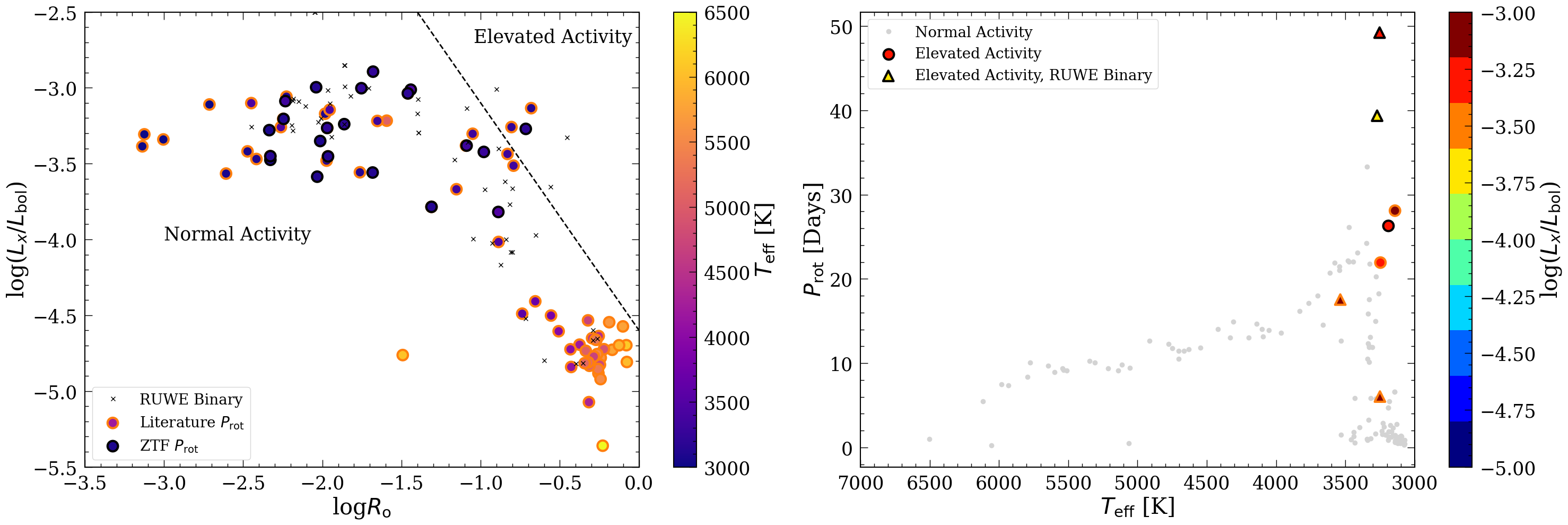}
\caption{The left panel shows $\log(L_x/L_{\rm bol})$ as a function of \ro, while the right panel shows the temperature--\prot\ diagram for stars with normal (gray points) and elevated activity (colored points), where elevated activity is defined by the dashed line in the left panel. 
Elevated activity is identified by the line crossing (log\ro, $\log(L_x/L_{\rm bol})$) = (-2, -1.6) and (0, -4.6), where the points lay above the general trend.
The elevated activity observed among the slowly rotating fully convective M dwarfs suggests that these stars may have experienced enhanced magnetic braking, supporting an enhanced-braking origin for their unusually long rotation periods.
However, the activity-\ro\ relation exhibit large scatter and the slowest rotating stars with elevated activity are binaries identified by RUWE, as a result, more data is needed to draw any firm conclusion.
\label{fig:9.1}}
\end{figure*}

In addition, we cross-matched our sample with the eROSITA Hyades X-ray sample \citep{eROSITA, Khamitov2025} and obtained X-ray observations for 122 sources in the Hyades. 
We used the X-ray-to-bolometric luminosity ratio, $L_x/L_{\rm bol}$, provided in their catalog and calculated the Rossby number using the prescription described above. 
\autoref{fig:9.1} shows the results. 
The left panel shows $\log(L_x/L_{\rm bol})$ as a function of \ro, while the right panel shows the temperature--\prot\ diagram for stars with normal (gray points) and elevated activity (colored points), where elevated activity is defined by the dashed line in the left panel. 
Note that temperatures below 3000 K are less reliable, as the photometric temperature relation is extrapolated beyond its calibrated temperature range.
The elevated activity observed among the slowly rotating fully convective M dwarfs suggests that these stars may have experienced enhanced magnetic braking, supporting an enhanced-braking origin for their unusually long rotation periods.
However, the slowest rotating stars are binaries identified by RUWE, as a result, more data is needed to draw any firm conclusion.

\section{Implications for stellar dynamos and gyrochronology of fully convective stars}\label{sec:dynamo}
The presence of slowly rotating stars in the fully convective regime has important implications for our understanding of angular momentum evolution and magnetic dynamos in low-mass stars. 
Interestingly, our search was not restricted to slow rotators within a specific temperature range, yet the slow rotators we identify are found exclusively in the fully convective regime.
This is an important confirmation of the validity of gyrochronology for GK stars: there is no evidence for a population of young slow rotators that would have misleading ages from gyrochronology.

Our detection of very slow rotators in the FC regime provides an additional observational constraint showing that whatever the underlying dynamo mechanism, it must be capable of producing sufficient magnetic braking to spin some fully convective stars down to periods exceeding 35 days within $\lesssim1$ Gyr while leaving the rest rapidly rotating. 
This is particularly striking given that the average rapid-rotation lifetime is estimated to be $\sim$1--4 Gyr for fully convective stars with masses of $0.1$--$0.3\,M_\odot$ \citep{Pass2024}.

Because FC stars lack a tachocline, which the shear layer between the radiative core and convective envelope thought to play a key role in the solar $\alpha\Omega$ dynamo, it remains an open question whether a tachocline is required to generate the large-scale magnetic fields necessary for efficient angular-momentum loss.
Numerous studies have compared PC and FC stars through magnetic activity indicators \citep[e.g.,][]{Wright2016, Wright2018, Newton2017}, magnetic field strengths \citep[e.g.,][]{Morin2010, See2025}, and stellar rotation \citep[e.g.,][]{Newton2016, Pass2022, Lu2024, Chiti2024}.
These studies reveal an apparent tension in magnetic braking across the fully convective boundary: FC stars follow a similar rotation--activity relation to PC stars, yet rotation and Zeeman-Doppler Imaging (ZDI) measurements suggest stronger large-scale magnetic fields and potentially more efficient angular momentum loss in FC stars.

Two possibilities may reconcile these observations. 
First, the Rossby number (a dimensionless measure of how rapidly a star rotates relative to the timescale of its convective motions), \ro:=\prot/\tauc, may not have the same physical meaning in FC stars. 
The convective turnover time \tauc\ is difficult to define when the entire stellar interior is convective, and different prescriptions have been explored empirically \citep[e.g.,][]{Wright2011, Gossage2025} and through stellar evolution models \citep[e.g.,][]{Chiti2024, See2025, Gossage2025}.
If PC and FC stars operate different dynamos, an $\alpha\Omega$ dynamo associated with tachocline shear in PC stars \citep[e.g.,][]{Parker1955, Moffatt1978, Ossendrijver2003} versus an $\alpha^2$ dynamo in FC stars \citep[e.g.,][]{Steenbeck1966, kuker1999, Chabrier2006, Browning2008}, then different Rossby-number prescriptions may be required. 
Second, the relationship between large- and small-scale magnetic fields may change across the fully convective boundary. 
While these components are generally correlated in PC stars \citep[e.g.,][]{Vidotto2014}, FC stars may produce stronger large-scale fields that dominate magnetic braking while maintaining similar small-scale fields responsible for chromospheric and coronal activity \citep[e.g.,][]{Browning2008, See2025}. 
This could explain why activity diagnostics such as X-ray and H$\alpha$ emission show similar behavior across the boundary, while rotation and ZDI measurements indicate stronger magnetic braking in FC stars.

A key implication of our result is that it helps bridge the previously apparent gap between the rotational distributions of young clusters and field fully convective stars. 
Field M dwarfs exhibit a broad range of rotation periods, including both rapidly and slowly rotating stars, indicating that substantial spin-down occurs over their lifetimes \citep{Newton2016}. 
In addition, kinematic properties and binary systems suggest that some field M dwarfs may already be slowly rotating at relatively young ages \citep{Pass2022, Lu2024}. 
In contrast, young clusters such as Praesepe and the Hyades have historically shown predominantly rapid rotation in the fully convective regime, with few or no stars occupying the long-period tail. 
This apparent difference has made it difficult to establish a continuous evolutionary sequence connecting the rapid rotation observed in young fully convective stars to the slow rotation seen among older field M dwarfs.

These stars also provide an important test of gyrochronology at low stellar masses. 
Most empirical gyrochronology relations are calibrated using partially convective stars and young clusters, while the fully convective regime remains comparatively poorly constrained. 
The presence of \prot$>35$ day fully convective stars at ages of only $\sim600$--800 Myr demonstrates that stars of similar mass and age can span a much broader range of rotation periods than previously established. 
This directly challenges a central assumption of gyrochronology: that stars of a given mass or temperature converge onto a sufficiently tight rotation sequence that rotation period provides a unique age. 
If these slow rotators retain memory of their initial conditions or follow a distinct rotational evolutionary pathway, then a single-valued period--age relation may not exist for fully convective stars, at least at these ages.

In interpreting these long periods, however, it is also useful to distinguish the evolution of rotation period from that of angular momentum itself.
\citet{Kounkel2022} found that empirical gyrochronology relations expressed in terms of angular momentum remain comparatively smooth and continuous into the M-dwarf regime, despite the much more complicated morphology of the corresponding rotation-period distribution; \citet{Kounkel2023} inferred angular momenta and spin-down torques directly for pre-main-sequence stars and found systematic angular-momentum loss already operating within the first $\sim$10 Myr. 
The very long periods identified here extend this empirical picture to much later ages near and across the fully convective boundary. 
It would therefore be particularly useful to determine whether the slow rotators identified here remain a distinct population in angular-momentum space, or instead join a more continuous $J$--mass sequence once changes in stellar moment of inertia associated with structural evolution are taken into account.

Finally, these results highlight the need for longer-baseline rotation surveys of young clusters and nearby field M dwarfs. 
The scarcity of long-period measurements is currently a major limitation in establishing the rotational sequence of fully convective stars. 
Expanding the sample of old and young fully convective stars with reliable rotation periods, combined with spectroscopic measurements of magnetic fields and stellar metallicities, will allow us to determine whether the slow-rotator population reflects a change in dynamo behavior, enhanced magnetic braking, or a more gradual continuation of the rotational evolution established in partially convective stars.

\section{Conclusion}
We present a systematic search for long period ratotors in fully convective stars in the Hyades and Praesepe clusters, combining rotation periods from the literature with new measurements from ZTF. 
Our main conclusions are:
\begin{itemize}
    \item We identify a population of slowly rotating stars near or within the fully convective regime in Praesepe and the Hyades. 
    In total, we identify 15 stars with \prot$>35$ days, including 5 in Praesepe and 10 in the Hyades. 
    Of these, 10 have very low probabilities of field contamination, and 7 also show no signs of binarity based on multi-epoch radial velocities and Gaia RUWE. 
    These stars demonstrate that fully convective stars can reach rotation periods of a month or more by ages of $\sim600$--800 Myr.
    \item Gyrochronology assumes that, after a brief initial phase, stars with the same rotation period at a given mass or temperature share the same age. 
    While our data are consistent with this assumption for solar-type stars, the presence of young, slowly rotating stars near the fully convective boundary calls into question whether such a unique period--age relation applies at the lowest stellar masses.
    \item The slow rotators provide evidence that magnetic braking can already be strong at young ages in these stars, consistent with the existence of slowly rotating fully convective stars in older field populations.
\end{itemize}

Expanding the sample of fully convective stars with reliable long-period measurements, particularly in young clusters with well-constrained ages and metallicities, will be critical for determining whether the slow-rotator population reflects a change in dynamo behavior, enhanced magnetic braking, or a distinct evolutionary pathway.

\begin{acknowledgments}
J.G.F-T gratefully acknowledges the support provided by ANID Fondecyt Regular No. 1260371.
MHP acknowldeges support from NASA grant 80NSSC24K0622 and NSF grant 2507891.

This work has made use of data from the European Space Agency (ESA) mission Gaia,\footnote{\url{https://www.cosmos.esa.int/gaia}} processed by the Gaia Data Processing and Analysis Consortium (DPAC).\footnote{\url{https://www.cosmos.esa.int/web/gaia/dpac/consortium}} 
Funding for the DPAC has been provided by national institutions, in particular the institutions participating in the Gaia Multilateral Agreement.
This research also made use of public auxiliary data provided by ESA/Gaia/DPAC/CU5 and prepared by Carine Babusiaux. 
This research has also made use of NASA's Astrophysics Data System.
Funding for the Sloan Digital Sky Survey V has been provided by the Alfred P. Sloan Foundation, the Heising-Simons Foundation, the National Science Foundation, and the Participating Institutions. SDSS acknowledges support and resources from the Center for High-Performance Computing at the University of Utah. SDSS telescopes are located at Apache Point Observatory, funded by the Astrophysical Research Consortium and operated by New Mexico State University, and at Las Campanas Observatory, operated by the Carnegie Institution for Science. The SDSS web site is \url{www.sdss.org}.

SDSS is managed by the Astrophysical Research Consortium for the Participating Institutions of the SDSS Collaboration, including the Carnegie Institution for Science, Chilean National Time Allocation Committee (CNTAC) ratified researchers, Caltech, the Gotham Participation Group, Harvard University, Heidelberg University, The Flatiron Institute, The Johns Hopkins University, L'Ecole polytechnique f\'{e}d\'{e}rale de Lausanne (EPFL), Leibniz-Institut f\"{u}r Astrophysik Potsdam (AIP), Max-Planck-Institut f\"{u}r Astronomie (MPIA Heidelberg), Max-Planck-Institut f\"{u}r Extraterrestrische Physik (MPE), Nanjing University, National Astronomical Observatories of China (NAOC), New Mexico State University, The Ohio State University, Pennsylvania State University, Smithsonian Astrophysical Observatory, Space Telescope Science Institute (STScI), the Stellar Astrophysics Participation Group, Universidad Nacional Aut\'{o}noma de M\'{e}xico, University of Arizona, University of Colorado Boulder, University of Illinois at Urbana-Champaign, University of Toronto, University of Utah, University of Virginia, Yale University, and Yunnan University.
\end{acknowledgments}

%

\vspace{5mm}
\facilities{Gaia \citep{gaia, Gaiadr3}, Kepler \citep{kepler}, K2 \citep{K2}, PO:1.2m \citep[ZTF][]{ztfdata, ztftime}, TESS \citep{tess}}


\software{\texttt{astropy} \citep{astropy:2013, astropy:2018, astropy2022}, \texttt{lightkurve} \citep{lightkurve}, \texttt{Matplotlib} \citep{matplotlib}, \texttt{NumPy} \citep{Numpy}, \texttt{Pandas} \citep{pandas}, \texttt{Scipy} \citep{SciPy}, \texttt{ChatGPT} \citep{openai2025chatgpt}, \texttt{GaiaUnlimited.selectionfunctions.binaries} \citep{CastroGindard2024},
\texttt{StellarEV$\_$emulator} (Branca et al. submitted.)}



\appendix

\section{Long Rotation Period Measurements ($>$ 35 Days)}\label{appx:longP}
\autoref{fig:A1} and \autoref{fig:A2} show the Lomb--Scargle (LS) periodograms and phase-folded light curves for the stars with $P_{\rm rot}>35$ days in the Hyades and Praesepe, respectively. Gaia DR3 661281206736808064 and Gaia DR3 661216949728995584 were also identified by \citet{Pass2022}, with ZTF rotation periods that agree with our measurements.
3 of the 15 long-period stars have K2SFF light curves (two in the Hyades and one in Praesepe), all of which show reasonable photometric modulation when folded at the ZTF-detected periods (third column in the figures). 
Long-period signals are more difficult to detect in K2 because of its shorter observing baseline. 
This is illustrated in the second column of the figures, where the folded light curves exhibit edge effects associated with the beginning or end of the observing campaign when folded at the periods identified from the K2 LS periodograms. 
Moreover, only one to two cycles are covered in the folded K2SFF light curves, further highlighting the importance of long-baseline observations for detecting long rotation periods.

For the remaining stars, only TESS light curves are available.
None show evidence for reliable short-period variability ($P_{\rm rot}<15$ days), as indicated by both the absence of strong peaks in the TESS LS periodograms and the low amplitudes of the folded light curves. 
Stars with genuinely short rotation periods are expected to exhibit photometric variability of order $\sim0.005$ \citep{Boyle2026}, substantially larger than the variations observed here. 
These results further suggest that the long periods identified by ZTF are unlikely to be aliases or spurious detections of short-period signals.

\begin{figure*}
    \centering
     \includegraphics[width=\textwidth]{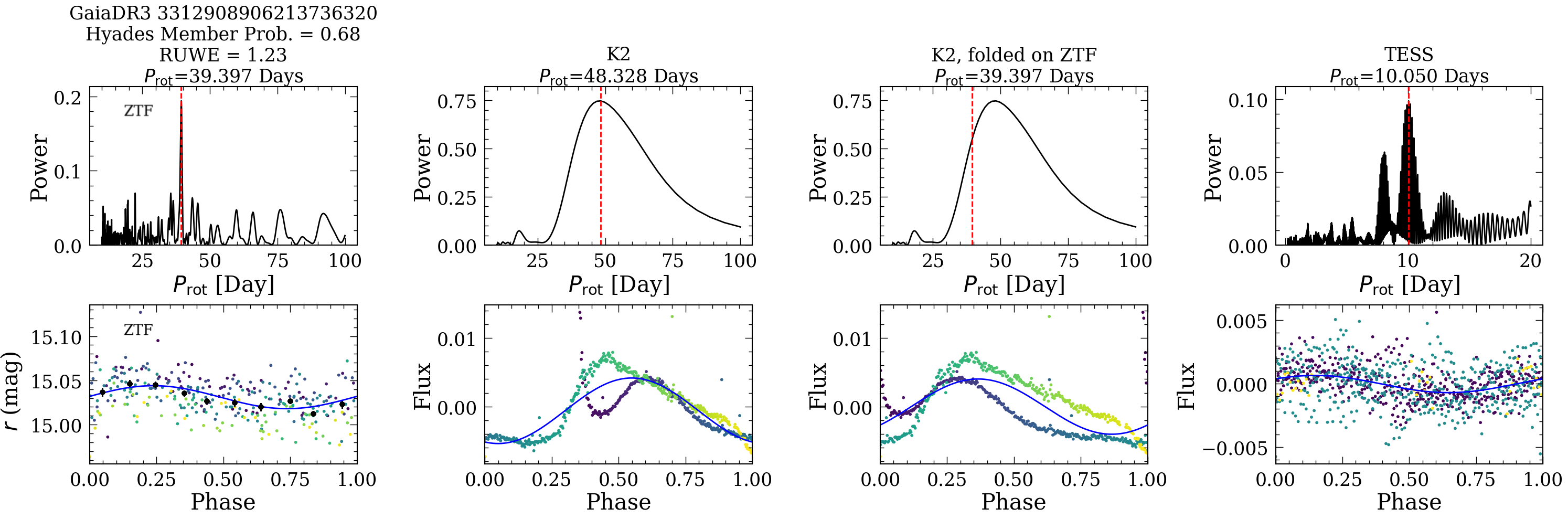}
     \includegraphics[width=\textwidth]{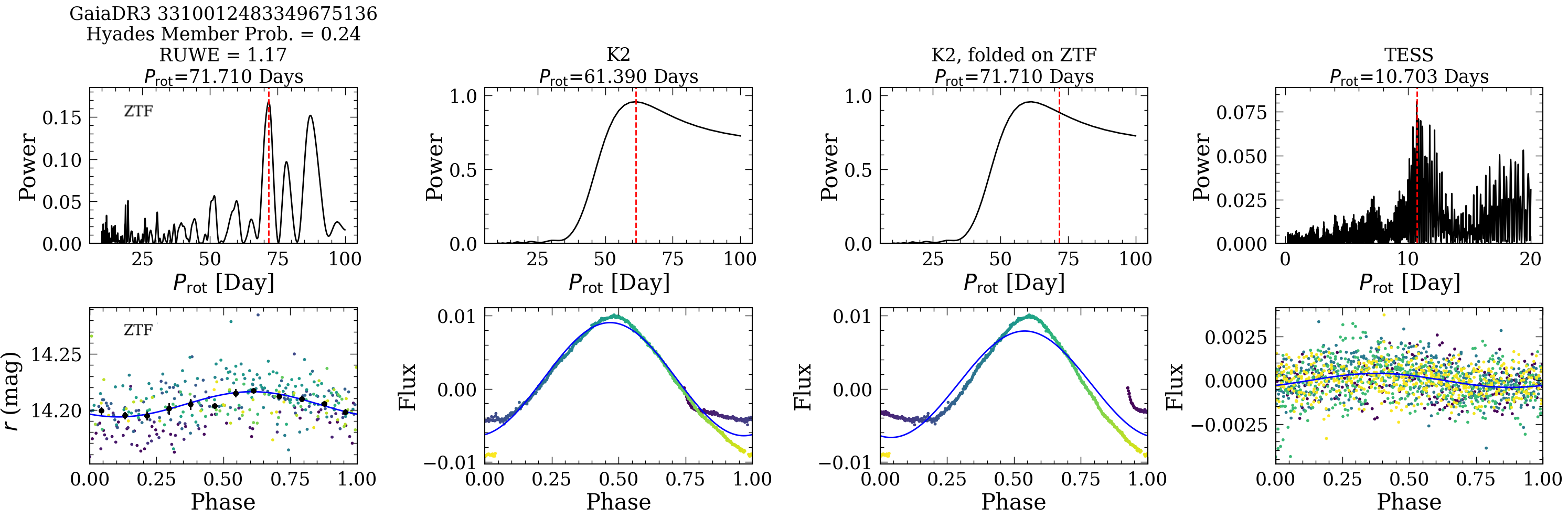}
    \includegraphics[width=0.49\columnwidth]{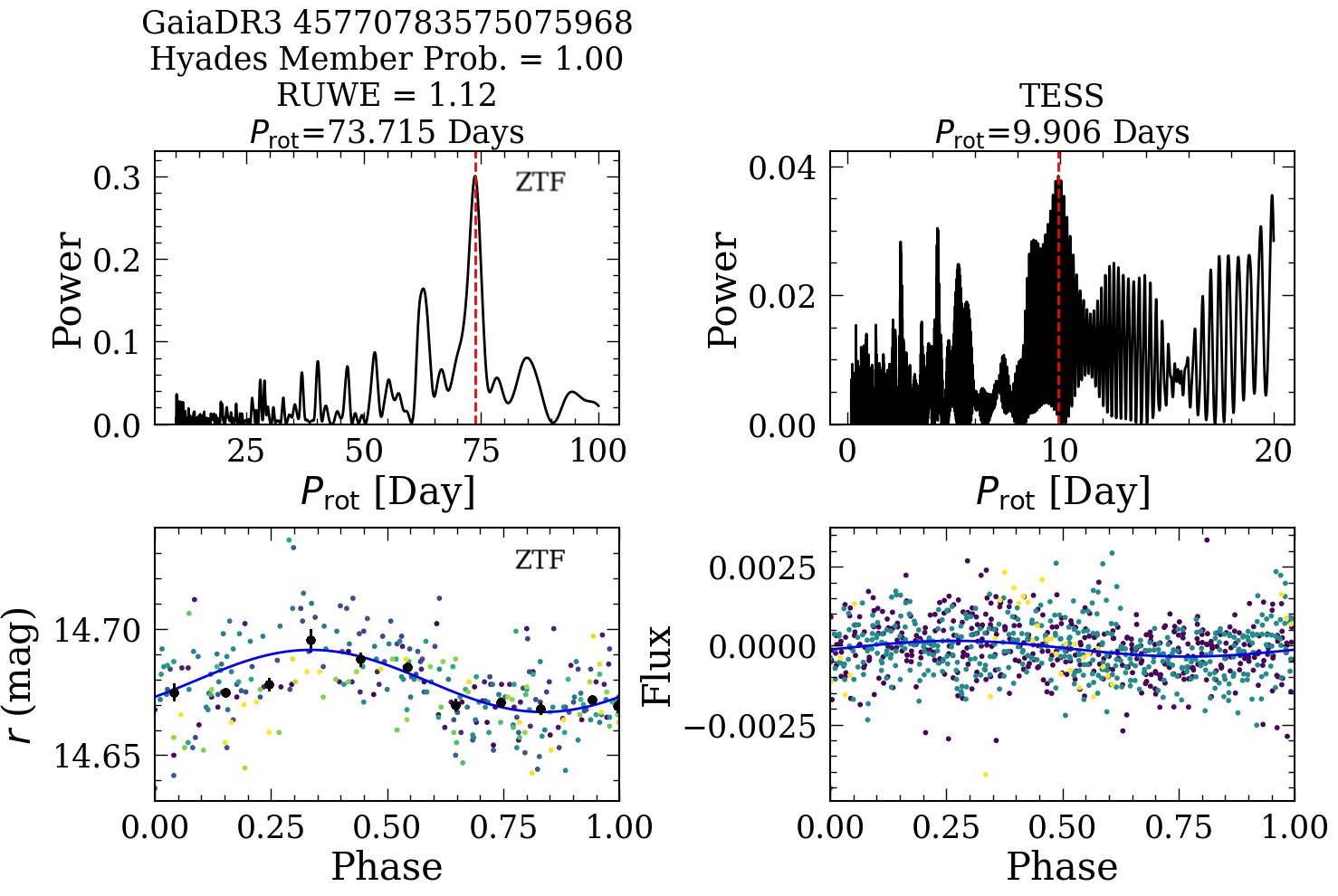}
    \includegraphics[width=0.49\columnwidth]{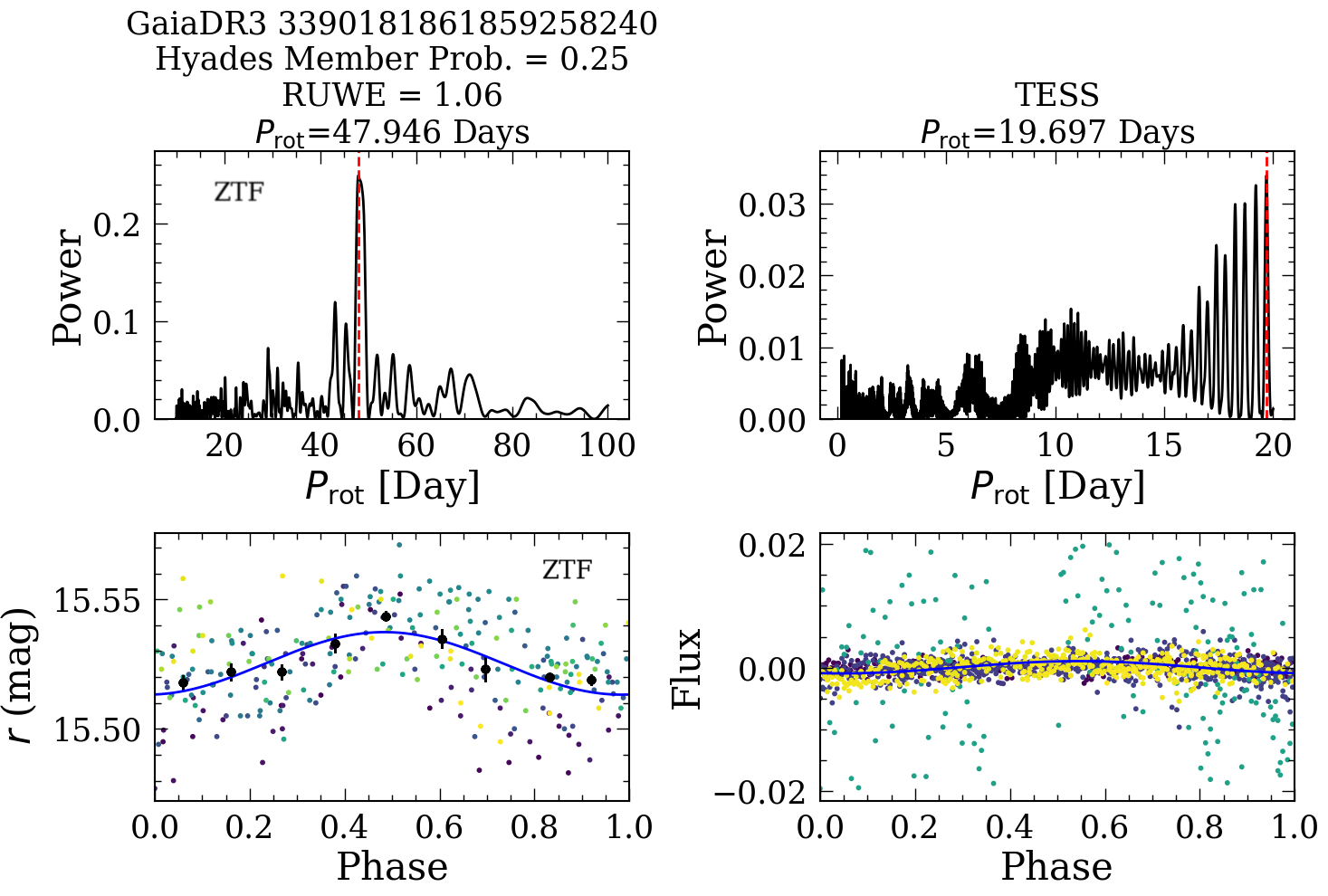}
    \caption{Similar to \autoref{fig:3}, but showing only the Lomb--Scargle periodograms and phase-folded light curves from the available surveys for Hyades stars with measured ZTF \prot$>35$ days.
    Only Gaia DR3 3312908906213736320 and Gaia DR3 3310012483349675136 have available K2SFF light curves.}
    \label{fig:A1}
\end{figure*}

\begin{figure*}
    \centering
    \addtocounter{figure}{-1}
     \includegraphics[width=0.49\columnwidth]{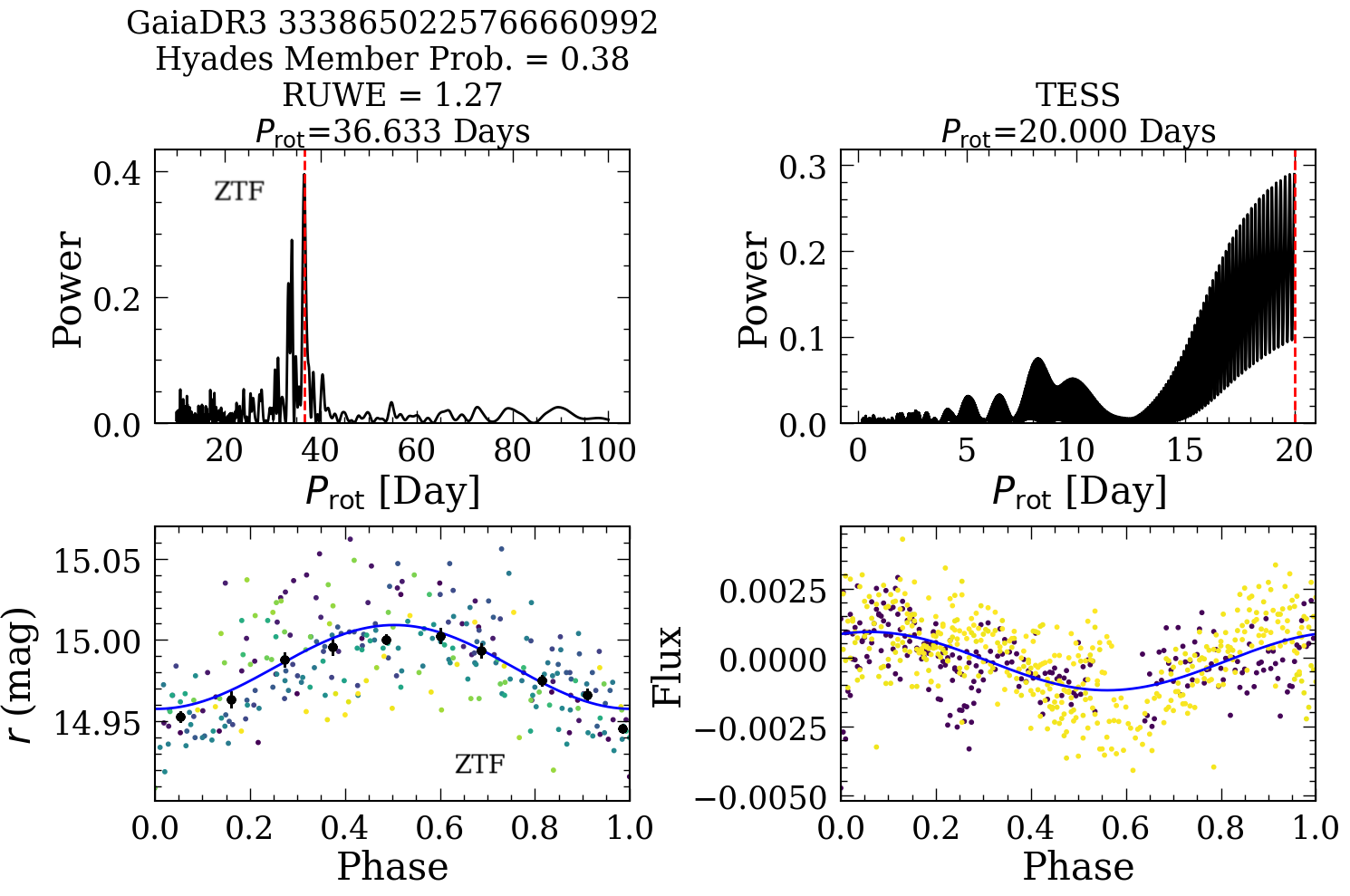}
    \includegraphics[width=0.49\columnwidth]{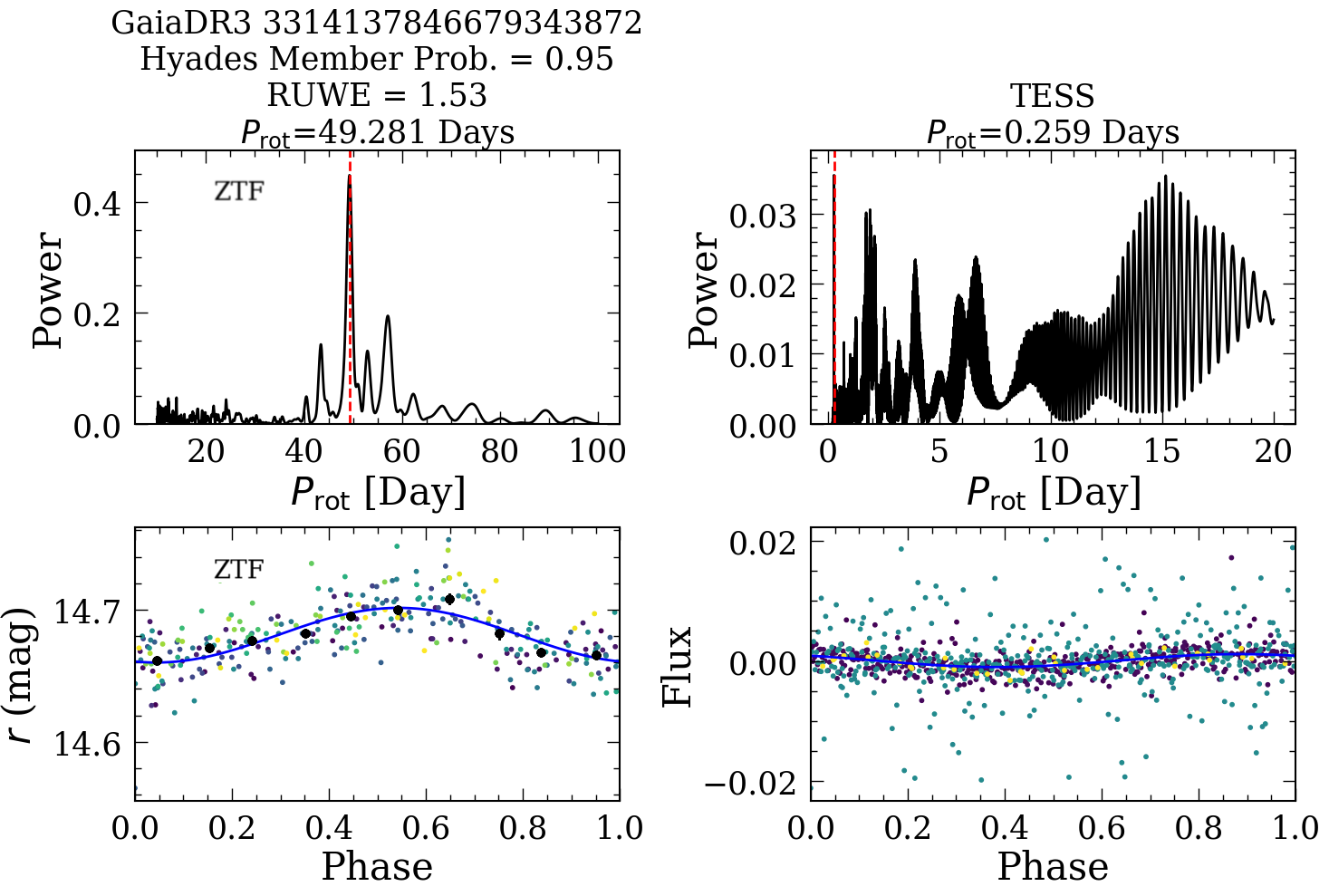}
     \includegraphics[width=0.49\columnwidth]{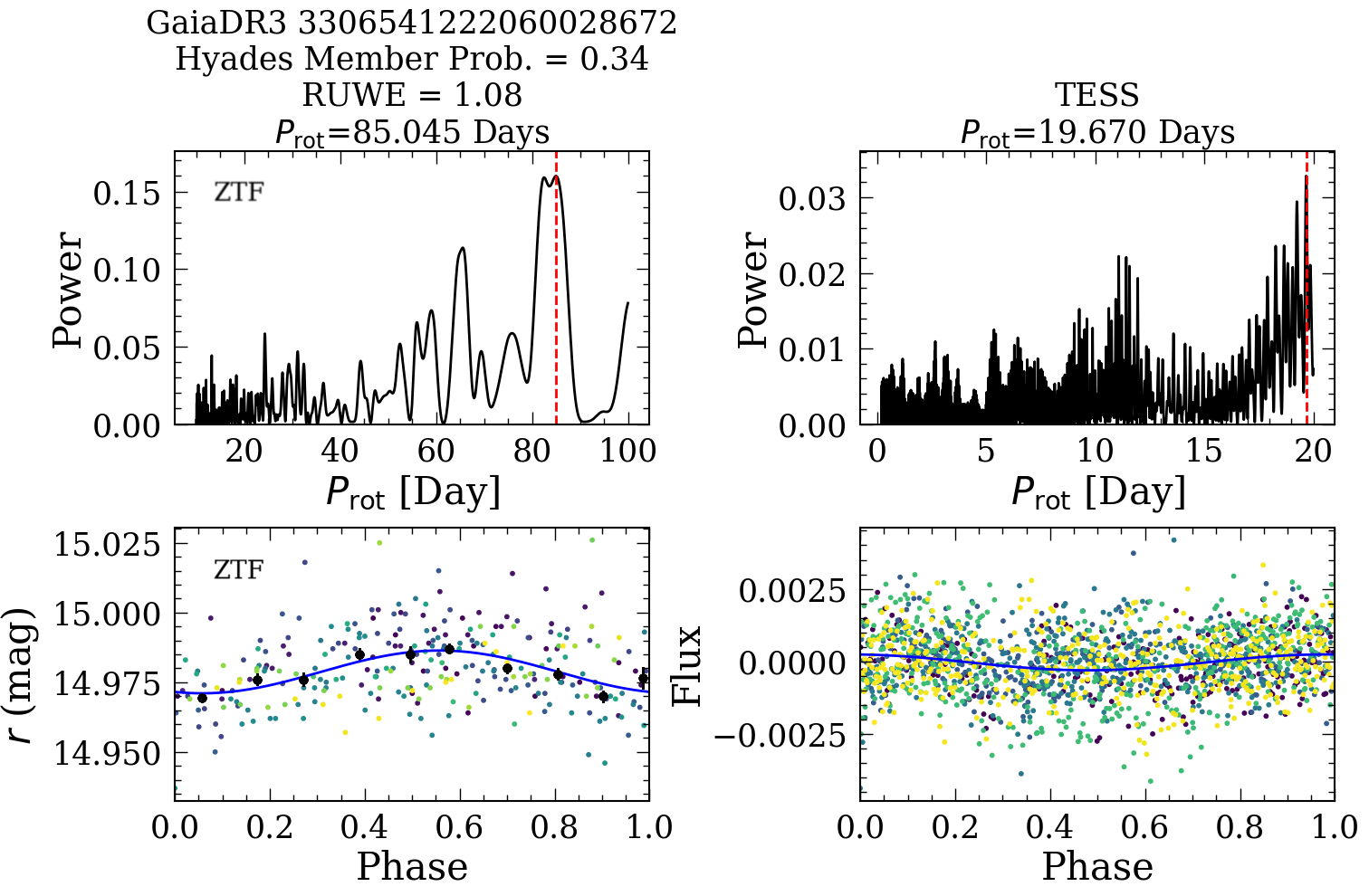}
     \includegraphics[width=0.49\columnwidth]{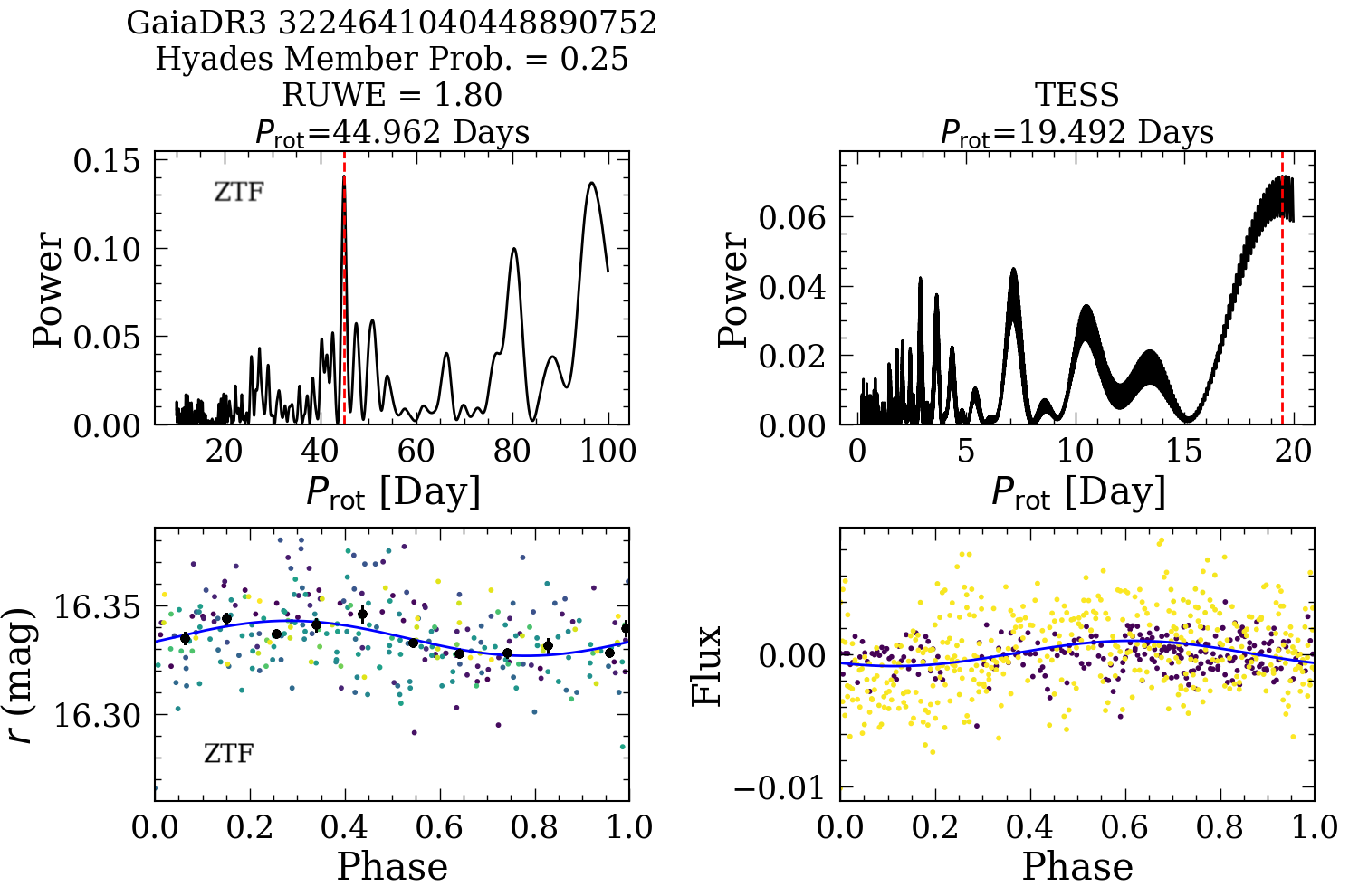}
      \includegraphics[width=0.49\columnwidth]{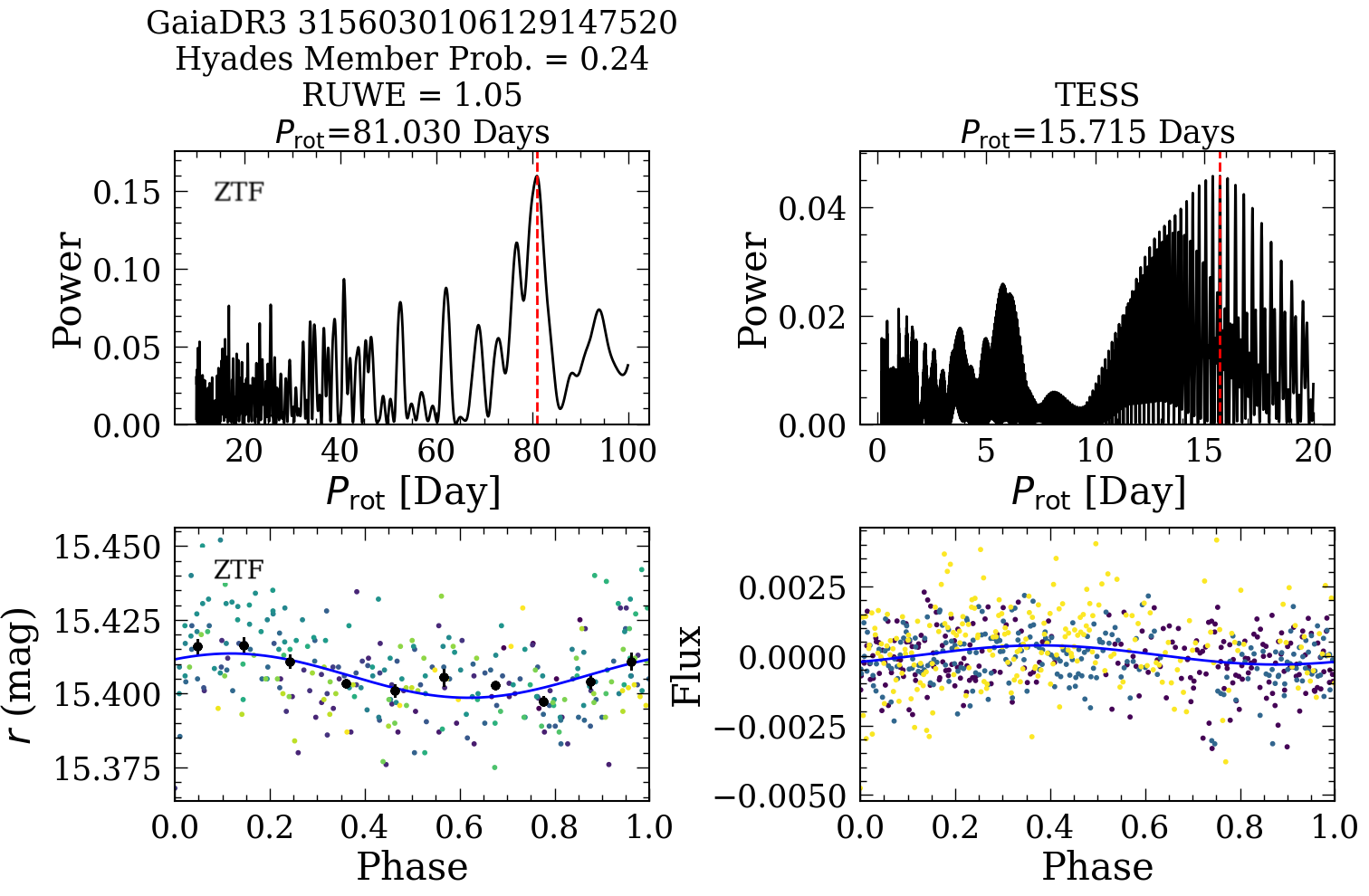}
      \includegraphics[width=0.49\columnwidth]{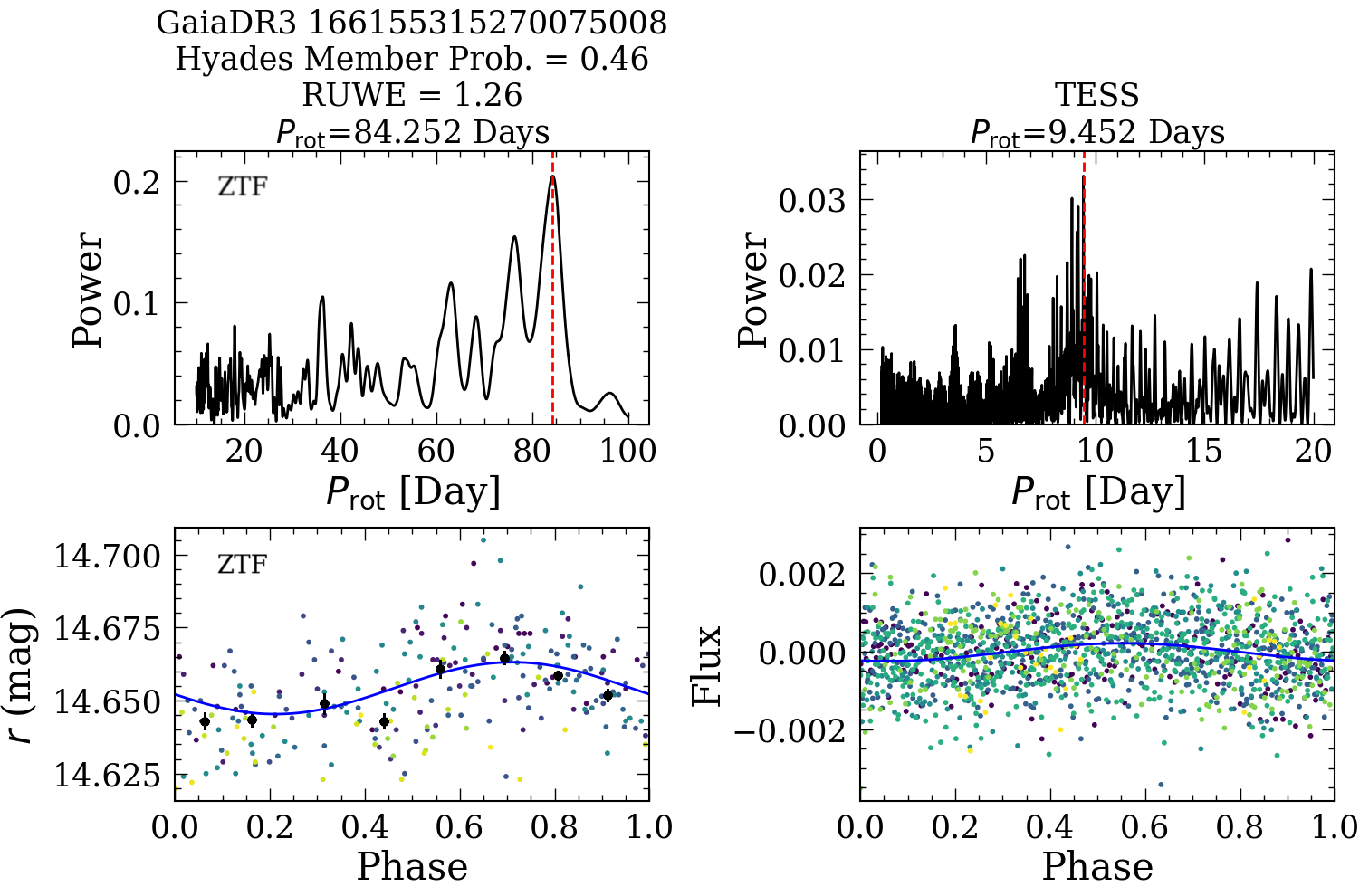}
      \caption{\autoref{fig:A1} continued.}
      \label{fig:A1_2}
\end{figure*}

\begin{figure*}
    \centering
    \includegraphics[width=\textwidth]{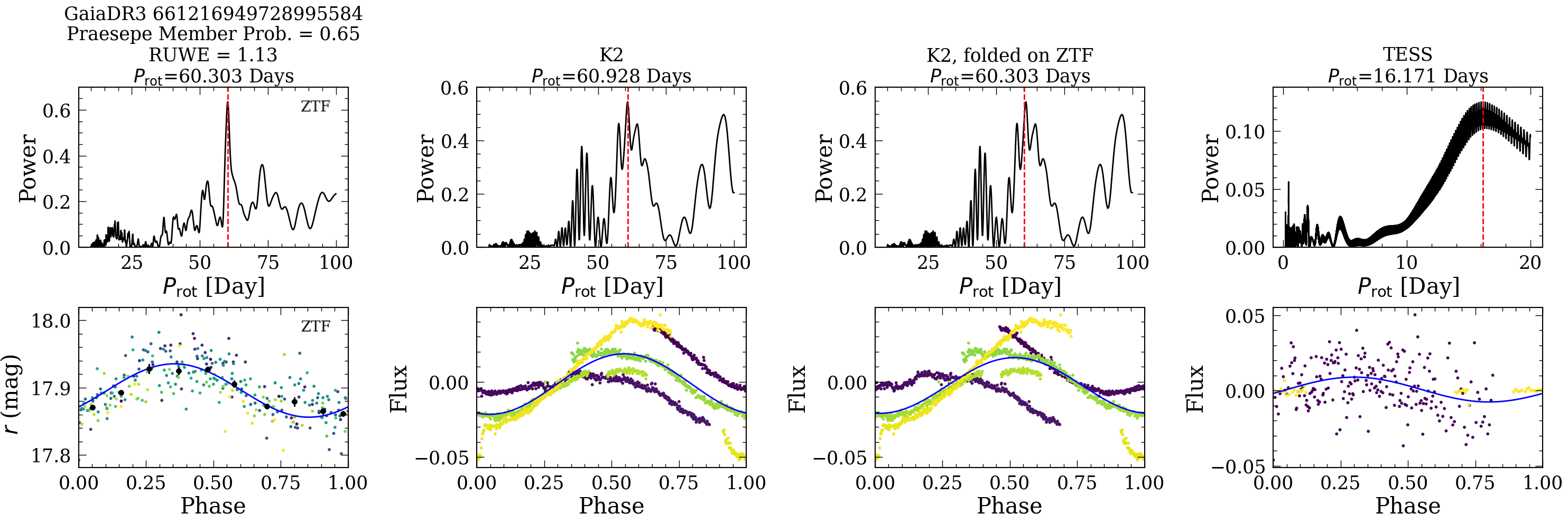}
    \includegraphics[width=0.49\columnwidth]{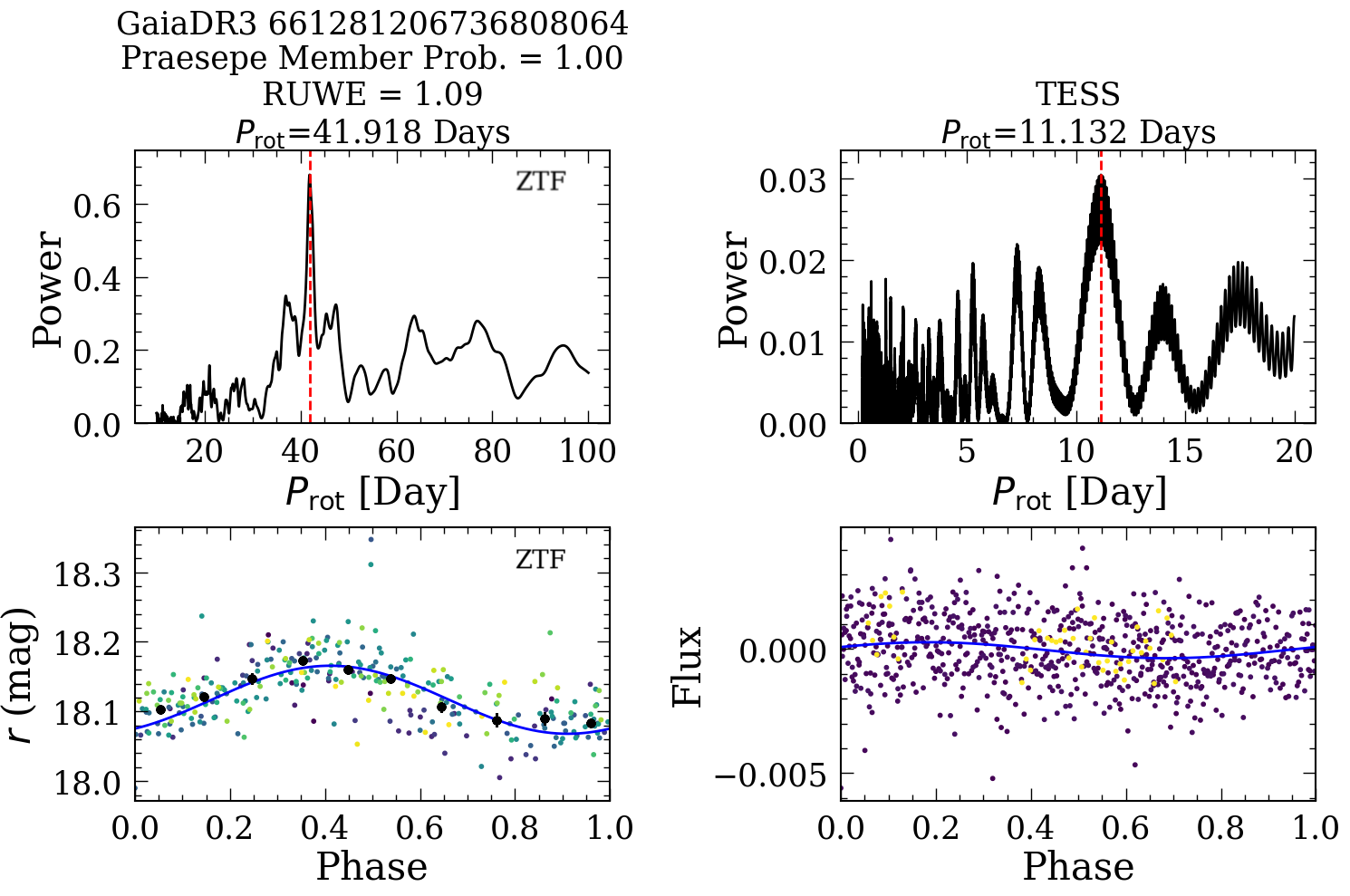}
    \includegraphics[width=0.49\columnwidth]{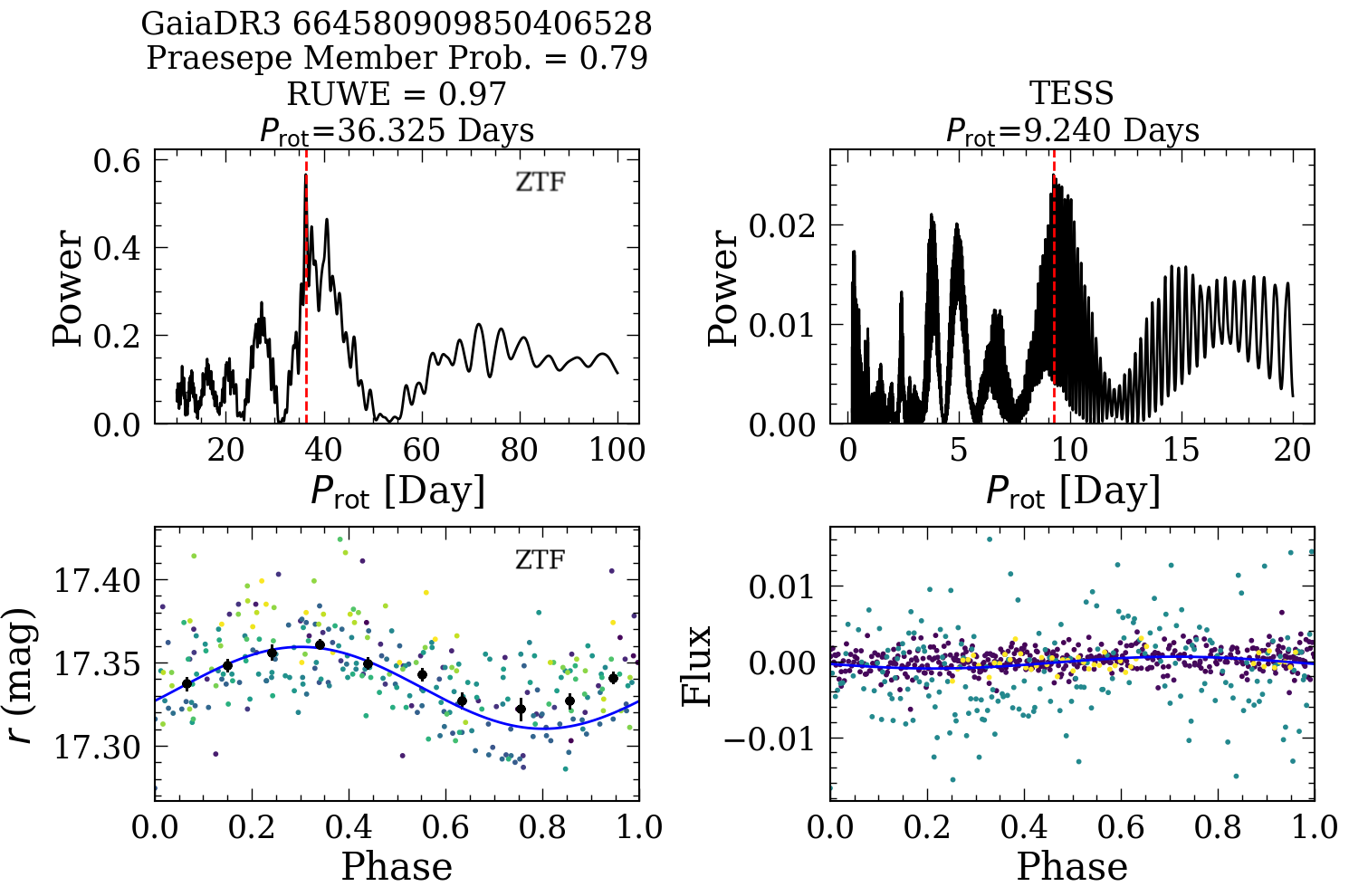}
     \includegraphics[width=0.49\columnwidth]{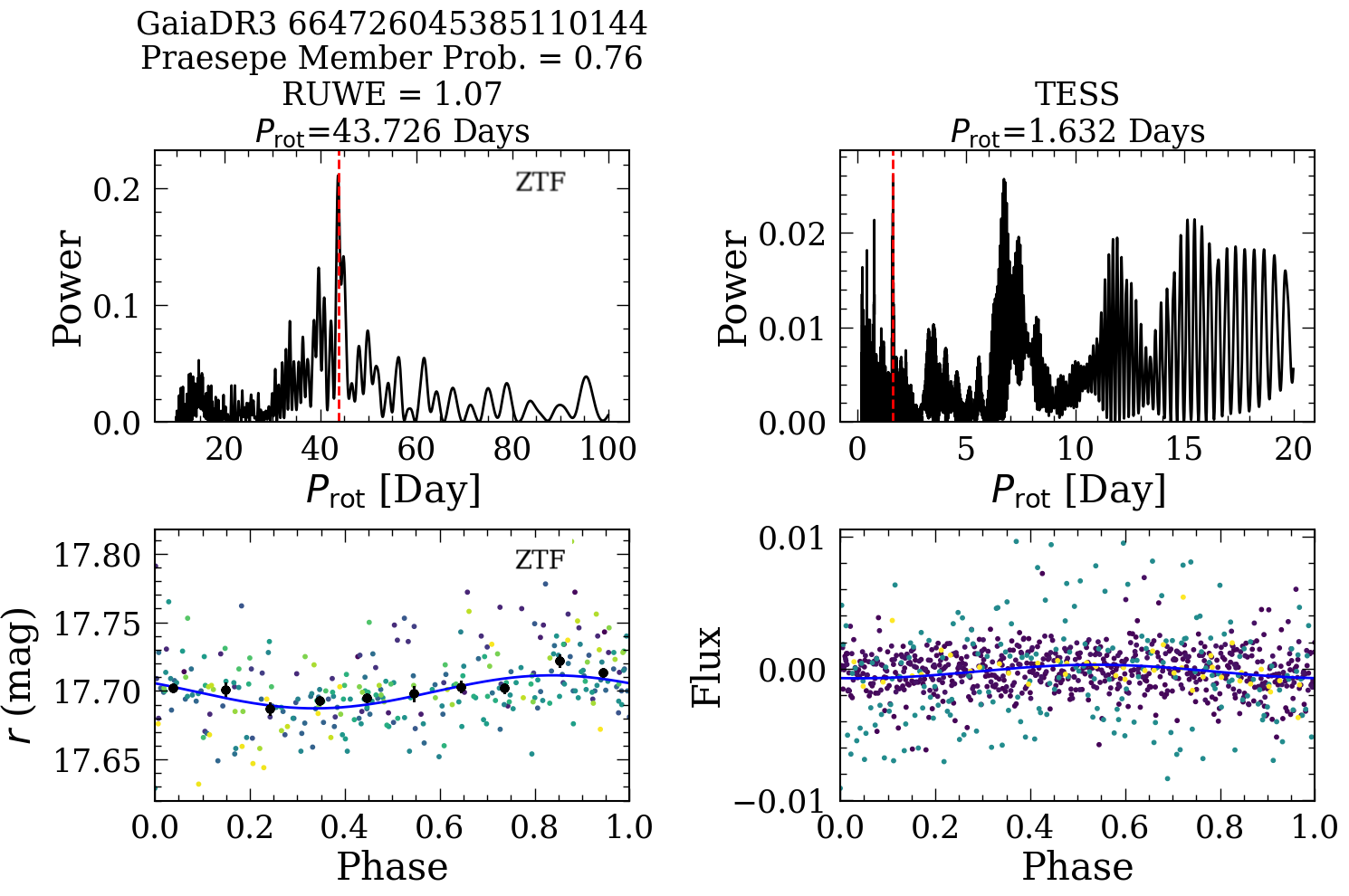}
    \caption{Same as \autoref{fig:A1} but for Praesepe.
    Only Gaia DR3 661216949728995584 has K2SFF light curve.
    Gaia DR3 661117134688210560 is shown in \autoref{fig:3}.
    \label{fig:A2}}
\end{figure*}

\bibliography{sample631}{}
\bibliographystyle{aasjournal}



\end{document}